\documentclass[aps,pra,twocolumn,letterpaper,superscriptaddress,10pt]{revtex4-2}
\usepackage{amssymb,amsthm,amsmath,amsfonts}
\usepackage{graphicx,ulem,enumerate,mathptmx}
\usepackage[pdftex,dvipsnames,usenames]{xcolor}
\usepackage[colorlinks=true,urlcolor=blue,citecolor=blue,linkcolor=blue]{hyperref}

\begin{document}

\title{Quantifying quantum coherence in polariton condensates}

\author{Carolin L\"uders}
    \email{carolin.lueders@tu-dortmund.de}
	\affiliation{Experimentelle Physik 2, Technische Universit\"at Dortmund, D-44221 Dortmund, Germany}

\author{Matthias Pukrop}
    \email{mpukrop@mail.uni-paderborn.de}
	\affiliation{Department of Physics and Center for Optoelectronics and Photonics Paderborn (CeOPP), Universit\"at Paderborn, 33098 Paderborn, Germany}

\author{Elena Rozas}
	\affiliation{Experimentelle Physik 2, Technische Universit\"at Dortmund, D-44221 Dortmund, Germany}

\author{Christian Schneider}
    \affiliation{Institute of Physics, University of Oldenburg, D-26129 Oldenburg, Germany}

\author{Sven H\"ofling}
    \affiliation{Technische Physik, Physikalisches Institut and W\"urzburg-Dresden Cluster of Excellence ct.qmat, Universit\"at W\"urzburg, 97074 W\"urzburg, Germany}

\author{Jan Sperling}
    \email{jan.sperling@upb.de}
    \affiliation{Integrated Quantum Optics Group, Institute for Photonic Quantum Systems (PhoQS), Paderborn University, Warburger Stra\ss{}e 100, 33098 Paderborn, Germany}

\author{Stefan Schumacher}
	\affiliation{Department of Physics and Center for Optoelectronics and Photonics Paderborn (CeOPP), Universit\"at Paderborn, 33098 Paderborn, Germany}
	\affiliation{Wyant College of Optical Sciences, University of Arizona, Tucson, Arizona 85721, USA}

\author{Marc A\ss{}mann}
	\affiliation{Experimentelle Physik 2, Technische Universit\"at Dortmund, D-44221 Dortmund, Germany}

\date{\today}

\begin{abstract}
	We theoretically and experimentally investigate quantum features of an interacting light-matter system from a multidisciplinary perspective, combining approaches from semiconductor physics, quantum optics, and quantum-information science.
	To this end, we quantify the amount of quantum coherence that results from the quantum superposition of Fock states, constituting a measure of the resourcefulness of the produced state for modern quantum protocols.
	This notion of quantum coherence from quantum-information theory is distinct from other quantifiers of nonclassicality that have previously been applied to condensed-matter systems.
	As an archetypal example of a hybrid light-matter interface, we study a polariton condensate and implement a numerical model to predict its properties.
	Our simulation is confirmed by our proof-of-concept experiment in which we measure and analyze the phase-space distributions of the emitted light.
	Specifically, we drive a polariton microcavity across the condensation threshold and observe the transition from an incoherent thermal state to a coherent state in the emission, thus confirming the buildup of quantum coherence in the condensate itself.
\end{abstract}

\maketitle

%%%%%%%%%%%%%%%%%%%%%%%%%%%%%%%%%%%%%%%%%%%%%%%%%%%%
%%%%%%%%%%%%%%%%%%%%%%%%%%%%%%%%%%%%%%%%%%%%%%%%%%%%
%%%%%%%%%%%%%%%%%%%%%%%%%%%%%%%%%%%%%%%%%%%%%%%%%%%%
\section{Introduction}

    Quantum physics is one of the most successful theories to describe nature.
    However, even during its conception, it was regarded as a counterintuitive formalism to interpret the physical world when compared to more familiar classical, thus accessible notions that do not involve quantum superpositions \cite{S35}.
    Therefore, various measures to quantify the incompatibility of quantum physics with classical concepts have been developed \cite{MBCPV12,SAP17,CG19}, initially being restricted to entanglement quantification tasks \cite{HHHH09}.
    Notably, recent approaches to quantifying such quantum superpositions are based on their usefulness in quantum-information science and technology \cite{BCP14,LM14,SV15,WY16}, advancing fundamental questions to applications, e.g., quantum teleportation \cite{BBCJPW93}, secure quantum key distribution \cite{BB84}, boson sampling \cite{AA13}, to name but a few.

\begin{figure}[b]
    \centering
    \includegraphics[width=\columnwidth]{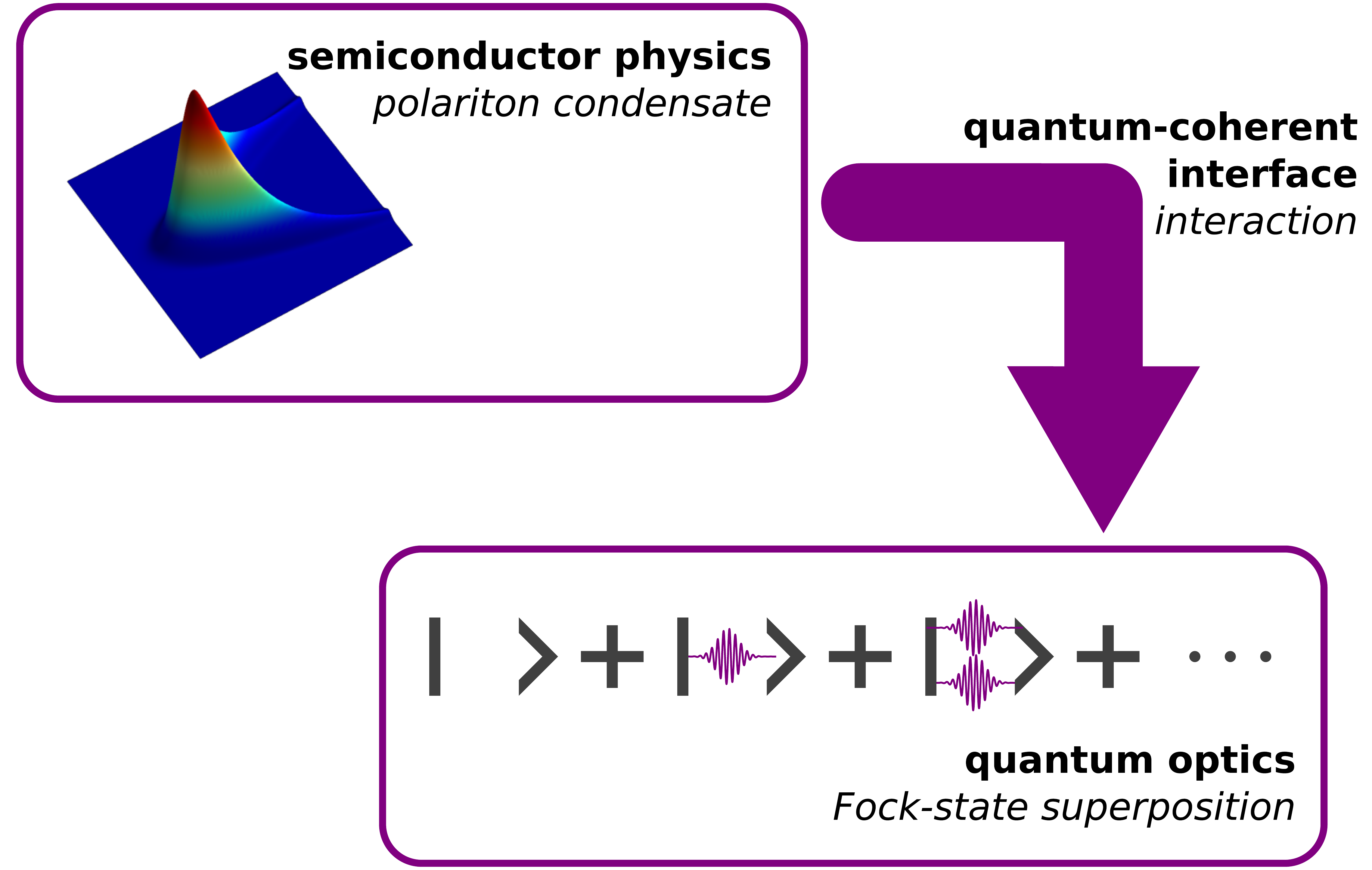}
    \caption{%
        Conceptual goal of our interdisciplinary research.
        A matter system represents a quantum device, e.g., a quantum processor, in which quantum superpositions are produced.
        A light field is used for distributing the generated quantum coherence, e.g., for quantum communications.
        The light-matter interaction serves as an interface.
        By analyzing the quantum features of the light field, quantum coherence of the semiconductor system can be characterized to assess the system's resourcefulness for quantum-information protocols.
    }\label{fig:interdis}
\end{figure}

    Along with the success of quantum theory, a branching out into many subfields occurred, each focusing on distinct systems and specific aspects thereof.
    For example, in today's research, semiconductor physics and quantum optics increasingly become separated fields, departing from their common origin.
    Thus, methods and findings from one branch of quantum physics are not always encountered in another subdomain, regardless of their general importance.
    Nonetheless, the emergence of quantum technologies \cite{NC00,DM03,D20} urges us to bring together different fields for developing interfaces between quantum processors, such as implemented in matter systems, and quantum-communication platforms, e.g., quantized radiation fields; see Fig. \ref{fig:interdis}.
    Here, we aim for combining contemporary concepts of semiconductor physics, quantum optics, and quantum-information science to study quantum coherence phenomena in both theory and experiment.

    The interplay of different quantum systems, such as the aforementioned light-matter systems, results in the notion of hybrid systems, which recently gained growing recognition because of the increasing demand for combining distinct technical implementations \cite{Aetal21}.
    In general, a hybrid quantum system consists at least of two distinct physical platforms (likewise, ways of physical information encoding) that are supposed to be able to carry the same quantum-information contents \cite{ANLF15,MHLJFL14,USPFL17}, thus being equipped with isomorphic quantum superposition structures \cite{SV15}.
    For example, a combination of a discrete-variable and a continuous-variable system can be combined to potentially teleport quantum superpositions between them, where one part is characterized by a discrete photon-number basis expansion and the other one by a continuous phase-space distribution \cite{USPFL17,ASCBZV17}.

    In the context of semiconductor quantum optics, first steps in the direction of implementing hybrid interfaces were made by theoretically analyzing \cite{KK06} and experimentally probing \cite{KKSHC11} the formation of excitons in semiconductors with quantum light.
    Promising candidates are exciton polaritons---referred to as polaritons hereafter---that describe light-matter quasiparticle excitations in semiconductor microcavities because of the strong coupling between excitons and photons \cite{KBML17}.
    Previous studies suggest that the quantum statistics of light leads to a nearly identical quantum statistics of the thereby generated quasiparticle matter state, whose properties are again mirrored by the emitted light.
    Additionally, nonlinearities of the polariton system may be exploited to create highly nontrivial states.
    For resonant excitation, quantum states of the emitted light, such as squeezed states \cite{BBA14}, have been demonstrated, and quantum correlations of the light source itself can be mapped onto the targeted matter system \cite{CMSVL15}.
    For nonresonant conditions, which we pursue in this work, polaritons form and relax towards the ground state of the system.
    A macroscopic population may form in this state which is accompanied by the spontaneous formation of coherence and off-diagonal, long-range order in the system \cite{LMKB04,KMSL07}.
    This state is a polariton condensate and may be considered as the driven-dissipative analog of a Bose-Einstein condensate.
    Furthermore, recent demonstrations generated photonic Fock states via polariton systems and characterized them \cite{DFSHSI19,MWJVBRLBANBRV19}.
    In one implementation of this concept, a polariton cavity was excited by a single photon that was part of an entangled photon pair, transferring the entanglement to the polaritons \cite{CCS18}.
    Still, the above works did not employ a unified concept of quantumness and focused on transferring quantum features of the prepared nonclassical light to the semiconductor system, not probing the coherence properties of the matter system itself via the emitted light as we are going to demonstrate here (Fig. \ref{fig:interdis}).

    While entanglement in hybrid systems is certainly useful \cite{KL12}, other forms of quantumness can exist beyond entanglement \cite{MBCPV12,ABC16,SAWV17,SPBW17}.
    For instance, a classical electromagnetic wave is most closely resembled by what is known as a coherent state, resulting in phase-space descriptions as introduced by Wigner, Glauber, and others \cite{S01,VW06}.
    And correlation functions can be used to discern classical wave and quantum particle properties, as observed for single photons \cite{HOM87}. 
    However, matter does require a different classical reference in which the wave properties are identified as quantum features and particles are the classical default, rendering commonly applied correlation-function-based and phase-space approaches to identify nonclassicality less useful.

    In this work, we specifically consider a superposition of particles to represent quantum phenomena in matter systems.
    This is in contrast to the standard concept of nonclassicality of light in which photons themselves, i.e., particles of quantized waves, are deemed highly nonclassical objects \cite{TEBSS21}.
    Thus, we probe for sophisticated particle quantum coherence effects within a polariton condensate via Fock-state superpositions in the emitted light field that carries the information about the matter system; see Fig. \ref{fig:interdis}.
    Thereby, the frequently held belief that only Fock states are resourceful is defied, which is additionally supported by applications, such as quantum random number generators \cite{SAL11,FXATW16} for quantum cryptography \cite{PABBBCWGLOPRSTUVVW20} and entanglement detection \cite{AMN07}, that can be carried out with coherent laser light.

    To partly overcome ambiguities of different classical notions \cite{P72}, the more agnostic concept of quantum coherence has been established, allowing for general classical references, while quantifying quantum superpositions on the merits of their resourcefulness for quantum-information applications; see Refs. \cite{SAP17,CG19} for recent reviews.
    Still, an open problem is to find a generally applicable and compatible concept of quantumness that applies to all distinct parts of a hybrid system in the same manner.
    Here, we are making a step forward to resolving this problem by analyzing discrete-variable particle quantum coherence through emitted, continuous-variable radiation fields.
    This is at the core of the challenging task to devise interfaces between different quantum technologies \cite{Aetal21}.
    Thereby, we also investigate the general quantum resourcefulness of states produced in a polariton system.

    In this contribution, we theoretically and experimentally analyze and quantify quantum coherence of a semiconductor system by monitoring the features of the emitted light.
    Specifically, we observe the transition from an incoherent thermal ensemble to a state that carries a significant amount of quantum coherence.
    (The quantum-information-based concept of quantum coherence must not be confused with the notion of macroscopic coherence that can have a classical or quantum origin and that is frequently used in semiconductor physics.)
    For quantifying quantum coherence, we determine the Hilbert-Schmidt distance between a generated state and its best incoherent approximation by reconstructing phase-space distributions.
    Although the general challenge of implementing a stable phase reference for the system under study is missing to date, our simulations enable us to analyze our data by filling in the blank.
    The combination of both numerical model of the nonlinear semiconductor system and quantum-optical, tomographic measurements thus enables us to implement a proof-of-concept demonstration in which we quantify quantum coherence that originates from quantum superpositions of particle states in a polariton condensate.

    The added value of investigating quantum coherence becomes clear when it is compared to other measures of coherence.
    Commonly studied coherence phenomena refer to correlations between fields at different points in time and space ($g^{(1)}$) and to correlations between intensities ($g^{(2)}$).
    While $g^{(1)}$ does not carry any information about whether the field under study is quantum or classical, $g^{(2)}$ and higher-order intensity correlation functions provide insight into the quantum phenomena of quantized radiation fields by means of their occupation number distribution. Accordingly, for a state with a given density matrix, $g^{(2)}$ depends only on the diagonal elements of the density matrix.
    By contrast, the concept of quantum coherence applied here provides insight into the quantum phenomena of a matter system by quantifying the superpositions of particles, which cannot interfere in classical matter and can be harnessed for quantum-information protocols.
    And quantum coherence depends only on the off-diagonal elements of the density matrix.
    Quantum coherence may thus be considered as complementary to earlier approaches in condensed-matter systems.
    While the buildup of macroscopic coherence and $g^{(2)}$ in polariton systems has been widely investigated \cite{LKW08,HSQHFY10,AAB15,KSF18}, there exist no studies on quantum coherence in condensed-matter systems to our knowledge.

    The remainder of this paper is structured as follows.
    In Sec. \ref{sec:QCoherence}, we establish the notion of quantum coherence, its immediate relation to the superposition principle, and its quantification using phase-space representations.
    A numerical simulation for the polariton system is carried out in Sec. \ref{sec:NumSim}, using a truncated Wigner approximation and Monte Carlo techniques.
    In Sec. \ref{sec:ExpRes}, the experiment is presented, and the results of the quantum coherence analysis from the optical system are discussed in Sec. \ref{sec:ResultsDiscuss}.
    We conclude in Sec. \ref{sec:Conclusion}.

%%%%%%%%%%%%%%%%%%%%%%%%%%%%%%%%%%%%%%%%%%%%%%%%%%%%
%%%%%%%%%%%%%%%%%%%%%%%%%%%%%%%%%%%%%%%%%%%%%%%%%%%%
%%%%%%%%%%%%%%%%%%%%%%%%%%%%%%%%%%%%%%%%%%%%%%%%%%%%
\section{Quantum-state characterization}\label{sec:QCoherence}

    In this section, we introduce the quantum-information side of our work.
    For this purpose, we consider different descriptions of quantum states and introduce the notion of quantum coherence.
    Specifically, we theoretically quantify the amount of superpositions of particle-number (likewise, Fock) states that are required to describe the bosonic system beyond commonly applied correlation-function-based techniques. 
    By employing light-matter interactions as an interface to optically probe quantum coherence in a polariton system, this approach provides the necessary framework to experimentally assess the quantum resources of the system under study.

%%%%%%%%%%%%%%%%%%%%%%%%%%%%%%%%%%%%%%%%%%%%%%%%%%%%
\subsection{Fock-basis expansion and quantum coherence}

    For simplicity, we begin with a single bosonic field, represented by the annihilation operator $\hat a$ and the creation operator $\hat a^\dag$.
    The state $\hat\rho$ of the system can be expanded via the eigenbasis of the number operator, $\hat n|n\rangle=n|n\rangle$ for $n\in\mathbb N$ and $\hat n =\hat a^\dag \hat a$.
    Recall that $n$ is the number of excitations of the given quantum field, i.e., the number of bosons.
    The desired state expansion reads
    \begin{equation}
    \label{eq:FockExpansion}
        \hat\rho=\sum_{m,n\in\mathbb N}\rho_{m,n}|m\rangle\langle n|.
    \end{equation}
    Herein, the diagonal elements $\rho_{n,n}=p_n$ correspond to classical probabilities, $p_n\geq0$ and $\sum_n p_n=1$.
    By contrast, off-diagonal elements, $\rho_{m,n}$ with $m\neq n$, are a result of quantum superpositions of basis states.

    It turns out that the off-diagonal entries can be used for implementing quantum protocols, being the defining foundation of the notion of quantum coherence \cite{SAP17,BCP14,LM14}.
    Conversely, an incoherent mixture of basis states,
    \begin{equation}
        \label{eq:IncState}
        \hat\rho_\mathrm{inc}=\sum_{n\in\mathbb N} p_n |n\rangle\langle n|,
    \end{equation}
    is considered a classical ensemble of particles.
    Whenever the state cannot be expressed in such a diagonal representation, $\hat\rho\neq\hat\rho_\mathrm{inc}$, the state of the system does exceed the notion of a classical ensemble and demands genuine quantum superpositions to be fully described.
    For example, a superposition state $|\Psi\rangle=\sum_{n}\psi_n|n\rangle$ with at least two nonzero expansion coefficients $\psi_{n_1}$ and $\psi_{n_2}$ results in a density operator $\hat\rho=|\Psi\rangle\langle\Psi|$ with a nonvanishing off-diagonal component $\rho_{n_1,n_2}=\psi_{n_1}\psi_{n_2}^\ast$, being a direct consequence of the superposition principle.
    It is worth mentioning that quantum coherence straightforwardly generalizes to multipartite systems in which incoherent mixtures of the states $|n_1\rangle\otimes|n_2\rangle\otimes\cdots$ form the classical reference.
    Furthermore, the observation that the binary representation of a single integer $n$ can be read as a string of zeros and ones might be helpful to understand the quantum-computational importance of quantum coherence in superposition states.

    A way to quantify the amount of quantum coherence is thus given by adding up the modulus square of the off-diagonal contributions \cite{BCP14} (see also Appendix \ref{app:QuantCoh} for details),
    \begin{equation}
    \label{eq:QCoherence}
    \begin{aligned}
        \mathcal C(\hat\rho)=&\sum_{m,n\in\mathbb N:m\neq n}|\rho_{m,n}|^2
        \\
        =&\|\mathrm{\hat\rho-\hat\rho_\mathrm{inc}}\|_\mathrm{HS}^2
        =\mathrm{tr}\left(\hat\rho^2\right)-\mathrm{tr}\left(\hat\rho_\mathrm{inc}^2\right),
    \end{aligned}
    \end{equation}
    being equivalently expressed as the distance of the state $\hat\rho$ and its best incoherent counterpart $\hat\rho_\mathrm{inc}$, having the same diagonal elements as the actual state \cite{SW18}.
    Here, the distance is given in the Hilbert-Schmidt norm, $\|\hat A\|_\mathrm{HS}=[\mathrm{tr}(\hat A^\dag\hat A)]^{1/2}$.
    If no quantum superpositions are involved, $\hat\rho=\hat\rho_\mathrm{inc}$, this measure of quantum coherence is zero.
    And it is strictly larger than zero for states that obey $\hat\rho\neq\hat\rho_\mathrm{inc}$, becoming larger with increasing off-diagonal contributions.

%%%%%%%%%%%%%%%%%%%%%%%%%%%%%%%%%%%%%%%%%%%%%%%%%%%%
\subsection{Phase-space representation for Fock-state-based coherence}

    Dissimilar to the discrete expansion of a quantum state in the Fock basis is the representation in terms of so-called coherent states, being a superposition of number-operator eigenstates, $|\alpha\rangle=e^{-|\alpha|^2/2}\sum_{n} \alpha^n |n\rangle/\sqrt{n!}$ for $\alpha\in\mathbb C$.
    For instance, this enables the expansion of the density matrix as
    \begin{equation}
        \label{eq:GSrep}
        \hat\rho=\int d^2\alpha\,P(\alpha)|\alpha\rangle\langle \alpha|,
    \end{equation}
    where $P$ refers to the Glauber-Sudarshan distribution \cite{G63,S63}.
    This phase-space distribution is often highly singular \cite{S16}, thus a convolution with a Gaussian kernel can be done to achieve a regularization, resulting in the prominent Wigner function $W(\alpha)$ \cite{W32} and Husimi function $Q(\alpha)$ \cite{H40}.
    Note that the latter phase-space distribution can be also obtained through the relation $Q(\alpha)=\langle\alpha|\hat\rho|\alpha\rangle/\pi$.
    See Refs. \cite{SW18,SV20} for comprehensive introductions to phase-space and so-called quasiprobability representations.

    Coherent states resemble classical oscillators most closely; thus they serve as the classical reference for waves but identify quantum particles as nonclassical \cite{TG65,M86}.
    Specifically, the inconsistency of $P$ with a non-negative probability distribution defines nonclassical radiation fields in quantum optics.
    Here, however, we want to identify particles as classical in order to provide a characterization of the matter system.
    Historically, matter has been understood as a collection of particles.
    Consequently, they here serve as the classical gauge for determining genuine quantum features in matter in the form of their superpositions.
    Thus, we ought to employ number states $|n\rangle$ as our classical reference [Eq. \eqref{eq:IncState}].
    Nevertheless, we can still express the coherence $\mathcal C(\hat\rho)$ and the corresponding and required incoherent state $\hat\rho_\mathrm{inc}$ in terms of continuous-variable phase-space representations (Appendix \ref{app:QuantCoh}), such as the Wigner, Husimi, and Glauber-Sudarshan function.

    In general, nonclassicality in the quantum-optical sense and quantum coherence in the number basis are complementary notions.
    That is, a Fock state $|n\rangle$ exhibits quantum signatures in the former sense but is classical in the latter context, and vice versa for a coherent state $|\alpha\rangle$.
    (In the discussion here, vacuum $n=0=\alpha$ is excluded in both cases for sake of exposition.)
    For instance, correlation-function-based nonclassicality criteria can detect that number states are, in fact, nonclassical;
    this means that no Glauber-Sudarshan function $P(\alpha)$ exists that is both a classical probability density and describes the Fock state according to Eq. \eqref{eq:GSrep}.
    Conversely, the Fock expansion in Eq. \eqref{eq:FockExpansion} requires elements beyond diagonal ones when representing coherent states;
    i.e., the probability mass function $p_n$ is insufficient to expand coherent states, signified by $\mathcal C>0$.
    See Ref. \cite{SW18} for a broader discussion of the interplay of probabilities and various coherence phenomena.
    The elementary examples above, in which different states are deemed quantum through different approaches, demonstrate the fundamental distinction between the common notion of nonclassicality and the operational concept of quantum coherence for quantum information.

%%%%%%%%%%%%%%%%%%%%%%%%%%%%%%%%%%%%%%%%%%%%%%%%%%%%
\subsection{Stimulated emission and displaced thermal states}

    As an example with relevance for our particular study, we may consider the essential model of a laser by identifying the single mode under study with an optical bosonic field.
    The lasing threshold is the lowest power applied to a nonlinear medium---being the polariton system in our case---at which a laser's output is dominated by stimulated (i.e., coherent) emission of excited particles rather than by spontaneous (i.e., incoherent, specifically thermal) emission.
    Thus, in transition between both regimes, the optical state is well approximated by a displaced thermal state for which the phase-space function takes the form \cite{VW06}
    \begin{equation}
        \label{eq:DefPfctDTS}
        P(\alpha)=\frac{\exp\left(-\frac{|\alpha-\alpha_0|^2}{\bar n}\right)}{\pi\bar n},
    \end{equation}
    defining a displaced thermal state with the coherent amplitude $\alpha_0\in\mathbb C$ and a width given by $\bar n>0$.
    We emphasize that the $P$ function in Eq. \eqref{eq:DefPfctDTS} is a non-negative distribution, hence not able to exhibit nonclassicality in the sense of quantized light  \cite{TG65,M86}.
    Yet, it can lead to quantum coherence as discussed in the following.

    Far below threshold, we expect a thermal emission for which $\alpha_0\to0$ holds true.
    Then, in Fock basis, we get the incoherent state $\hat\rho=(\bar n+1)^{-1}\sum_n(\bar n/[\bar n+1])^n |n\rangle\langle n|$, with a mean thermal photon number $\bar n=\langle\hat n\rangle$.
    When coherence is the dominating factor, $\bar n\to0$, we obtain the coherent state $\hat\rho=|\alpha_0\rangle\langle\alpha_0|$ with off-diagonal elements of the form $\rho_{m,n}=e^{-|\alpha_0|^2}\alpha_0^m\alpha_0^{\ast n}/\sqrt{m!n!}\neq0$ and a coherent photon number $|\alpha_0|^2=\langle\hat n\rangle$.
    Consequently, this state exhibits a high amount of coherence.
    In fact, the laser, producing coherent light, has been one of the first commercial applications of such quantum superpositions.
    It is further worth pointing out that a general relation exists between laserlike processes and the presence of general notions of quantum coherence \cite{A14}.

\begin{figure}[b]
    \centering
    \includegraphics[width=.7\columnwidth]{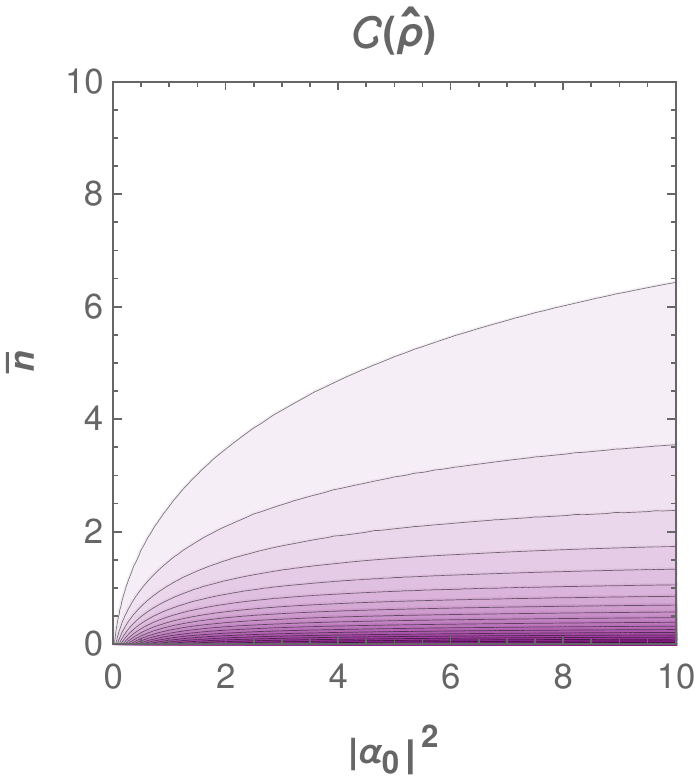}
    \includegraphics[width=.15\columnwidth]{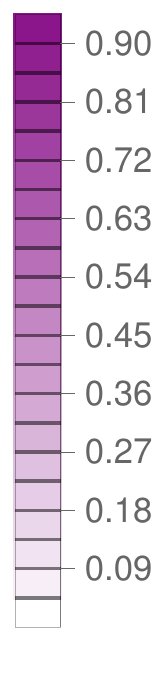}
    \caption{%
        Quantum coherence $\mathcal C(\hat\rho)$ as a function of the coherent and thermal photon numbers, $0\leq|\alpha_0|^2\leq 10$ and $0\leq \bar n\leq 10$, respectively.
        With increasing displacement $\alpha_0$, the amount of quantum coherence increases.
        And a reduction of coherence is observed when the thermal background $\bar n$ increases.
    }\label{fig:QuantCohDTS}
\end{figure}

    In our case, the coherence measure from Eq. \eqref{eq:QCoherence} for displaced thermal states takes the form
    \begin{equation}
    \label{eq:Coherence}
		\mathcal C(\hat\rho)=\frac{
			1-
			\exp\left[-\frac{2|\alpha_0|^2}{(\bar n+1)^2-\bar n^2}\right]
			I_0\left[\frac{2|\alpha_0|^2}{(\bar n+1)^2-\bar n^2}\right]
		}{(\bar n+1)^2-\bar n^2},
	\end{equation}
	where $I_0$ is the zeroth modified Bessel function of the first kind.
    See Appendix \ref{app:DisThState} for technical details.
    As one expects, the coherent photon number $|\alpha_0|^2$ ought to be high while the thermal photon number $\bar n$ ought to be low for producing a high amount of quantum coherence, as shown in Fig. \ref{fig:QuantCohDTS}.
    In fact, $\mathcal C$ is a monotonically increasing function with the coherent amplitude $|\alpha_0|$ and monotonically decreasing with $\bar n$ (Appendix \ref{app:EntanglementMonotonocity}).
    In the limit of a large coherent amplitude, $|\alpha_0|\to\infty$, and vanishing thermal background, $\bar n\to 0$, the quantum coherence saturates at a maximum value of 1.

    In the context of quantum technology applications, the predicted quantum coherence quantifies the states' usefulness in quantum protocols, which can exploit coherent states as a resource in different protocols \cite{SAL11,FXATW16,PABBBCWGLOPRSTUVVW20,AMN07} although they are deemed classical in the context of quantum optics.
    Moreover, a direct conversion of the particle-based quantum coherence to entanglement is possible \cite{KSP16}.
    For example, an incoherent pair-generation process that decays $n$ particles into $n$ pairs $|n\rangle\otimes|n\rangle$ can be considered; see Appendix \ref{app:EntanglementMonotonocity} for details.
    Such an incoherent operation cannot introduce coherence, but it directly transforms input coherence $\sum_n\psi_n|n\rangle$ into output entanglement $\sum_n \psi_n|n\rangle\otimes|n\rangle$ \cite{KSP16}, which in itself is a key resource for quantum-communication applications \cite{HHHH09}.

    Our goal is to quantify the quantum-information-based amount of quantum coherence of the emitted quantum light from a polariton condensate.
    Specifically, assuming that a linear interaction approximates the interaction between polaritons and emitted photons sufficiently well, the resulting emission resembles up to a rescaling the internal state of the condensate; see Appendix \ref{app:LinCoupling} for a proof. 
    Thus, the quantum coherence of the measured light field can serve as an interface that alludes to the quantum characteristics within the polariton condensate.

%%%%%%%%%%%%%%%%%%%%%%%%%%%%%%%%%%%%%%%%%%%%%%%%%%%%
%%%%%%%%%%%%%%%%%%%%%%%%%%%%%%%%%%%%%%%%%%%%%%%%%%%%
%%%%%%%%%%%%%%%%%%%%%%%%%%%%%%%%%%%%%%%%%%%%%%%%%%%%
\section{Simulation approach}\label{sec:NumSim}

\begin{figure}[t]
	\centering
 	\includegraphics[width=0.48\textwidth]{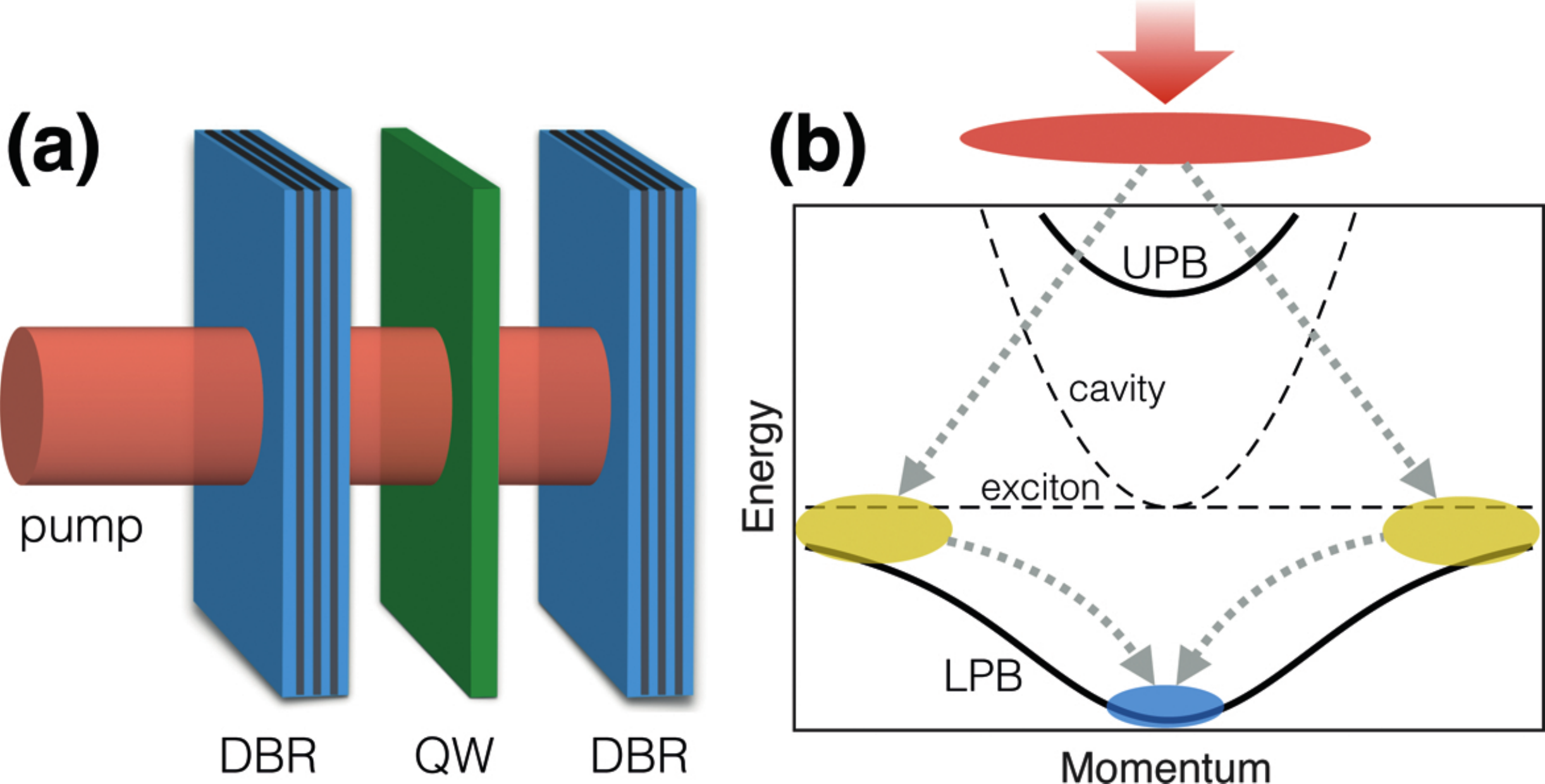}
 	\caption{%
        (a) Sketch of a semiconductor microcavity consisting of a quantum well (QW) that is embedded inside a cavity, realized by distributed Bragg reflectors (DBR).
        (b) Upper and lower polariton branch (UPB and LPB, respectively) alongside the bare cavity and exciton dispersions.
        A nonresonant pump injects free carriers (top, red) which form excitons and relax down to the LPB, building up a reservoir (left and right, yellow).
        From the reservoir, stimulated scattering results in the formation of the polariton condensate (bottom, blue) near the zero-momentum ground state.
    }\label{fig:sim0}
\end{figure}

    In the remainder of this contribution, we investigate polariton condensates that are nonresonantly excited in a planar semiconductor microcavity \cite{KBML17}.
    Therein, polaritons are described as hybrid light-matter quasiparticles resulting from the strong coupling between cavity photons and quantum-well excitons \cite{WNIA92}. 
    The strong coupling is characterized by a normal-mode splitting of the dispersion relation into an upper and lower polariton branch; see Fig. \ref{fig:sim0}(a). 
    Polaritons exhibit strong nonlinear interactions through their excitonic part while optical excitation and signal readout is mediated via the photonic part.

    A prominent and well-studied phenomenon is the condensation of polaritons into a macroscopic coherent state.
    An overview can be found in Refs. \cite{DHY10,CC12}.
    A nonresonant pumping beam far above the exciton resonance is used to inject free carriers into the system, forming excitons which relax down to the lower polariton branch and build up a reservoir, as outlined in Fig. \ref{fig:sim0}(b).
    Stimulated scattering into the zero-momentum ground state then creates a macroscopic condensate.
    Owing to the finite lifetime of polaritons, the system requires constant pumping in order to maintain the condensed state.
    And the latter is intrinsically out of equilibrium, with its steady state being in balance between loss and gain.

    In this section, we introduce a numerical model that implements the spatially resolved nonequilibrium condensation process, and we apply this model to investigate features of the ground state in momentum space.
    In particular, we study phase and intensity fluctuations of the polariton condensate and eventually determine the amount of quantum coherence present within the ground-state condensate by applying the concepts established in the previous section.
    In addition, these numerical results can predict the general features of the emitted light, as observed in our experiment, requiring only a few experimentally informed system parameters as input.

%%%%%%%%%%%%%%%%%%%%%%%%%%%%%%%%%%%%%%%%%%%%%%%%%%%%
\subsection{Model for numerical simulations}

    To describe the dynamics of the polariton condensate, we use an open-dissipative Gross-Pitaevskii equation coupled to an incoherent reservoir.
    Classical and quantum fluctuations are taken into account within a truncated Wigner approximation (TWA).
    Expectation values, specifically, coherence properties, can be calculated using Monte Carlo techniques.
    See Refs. \cite{SLC02,WS09,CC12} for comprehensive introductions.

    The main idea is to employ the Wigner representation for a bosonic polariton field operator $\hat{\psi}(\mathbf{r})$ that can be used to sample the phase-space distribution $W[\psi (\mathbf{r})]$.
    Therein, $\psi\in\mathbb C$ is the complex amplitude that describes the polariton field, analogously to the optical mode $\hat a$ and coherent amplitude $\alpha$ considered in the previous section.
    In particular, the TWA allows for mapping the time evolution of the Wigner distribution onto a set of stochastic partial differential equations for the corresponding phase-space variables.
    On a finite two-dimensional spatial $N\times N$ grid of length $L$, where $\mathbf{r}\equiv\mathbf{r}_i$, $i\in\lbrace 1,...,N^2\rbrace$ denotes a discrete grid point, the dynamics of the polariton field $\psi(\mathbf{r})$ when coupled to the incoherent reservoir density $n_\mathrm{res}(\mathbf{r})$ reads \cite{WS09}
    \begin{align}
        \nonumber
        d\psi=& -\frac{i}{\hbar}\bigg[H+\frac{\mathrm{i}\hbar}{2}\left( Rn_\mathrm{res}-\gamma_c \right)+g_r n_\mathrm{res}+g_c|\psi|_-^2 \bigg]\psi dt
        \\
        \label{eq:GPE_psi}
        &+dW,
        \\
        \label{eq:GPE_n}
        \frac{\partial n_\mathrm{res}}{\partial t}=&\left(-\gamma_r-R|\psi|_-^2 \right)n_\mathrm{res}+P\,.
    \end{align}
    Here $H=-\hbar^2\nabla^2/(2m_{\mathrm{eff}})$ is the free-particle Hamiltonian with the effective polariton mass $m_{\mathrm{eff}}$,
    $\gamma_c$ and $\gamma_r$ are the decay rates of the condensate and the reservoir,
    and the interaction strength between polaritons is described by $g_c$.
    The condensation rate and condensate-reservoir interaction are given by $R$ and $g_r$.
    Furthermore, the renormalized condensate density reads $|\psi|^2_-{\equiv}|\psi|^2{-}(\Delta V)^{-1}$, where $\Delta V=L^2/N^2$ denotes the unit-cell volume of the two-dimensional grid.
    The, in general, complex Wiener noise contribution is given by $dW$, with correlations satisfying
    \begin{align}
    \begin{aligned}
        \langle dW(\mathbf{r})dW(\mathbf{r'}) \rangle &=0
        \\
        \text{and}\quad
        \langle dW(\mathbf{r})dW^\ast(\mathbf{r'}) \rangle
        &=\left( Rn_\mathrm{res}+\gamma_c \right)\frac{\delta_{\mathbf{r},\mathbf{r'}}\,dt}{2\Delta V}.
    \end{aligned}
    \end{align}
    Expectation values of arbitrary products of the field operators (in symmetric ordering as implied by the Wigner function \cite{VW06}) can be calculated as the average over many stochastic realizations.
    Additional details on the numerical implementation can be found in Appendix \ref{app:NumDetails}.

    The TWA method was successfully utilized to describe properties of different types of Bose gases beyond the mean-field approximation, e.g., ferromagnetic domain formation in spinor Bose-Einstein condensates \cite{SLSC09}, quantum correlations of signal-idler emission in the ring-shaped luminescence of resonantly excited semiconductor microcavities \cite{VCC07}, and analogs of black-hole Hawking radiation processes in a polariton fluid \cite{GC12}. 
    Recently, the TWA method has been also used to study critical exponents and the Kibble-Zurek mechanism in polariton condensates \cite{CDZCPS18,ZDCCPS20}.

%%%%%%%%%%%%%%%%%%%%%%%%%%%%%%%%%%%%%%%%%%%%%%%%%%%%
\subsection{Numerical Results}

    For applying the previously introduced approach to simulate our system, we suppose a continuous-wave pump with super-Gaussian profile $P(\mathbf{r})=P_0\exp\left[-\mathbf{r}^4/w^4\right]$, with width $w=65~\mathrm{\mu m}$.
    It is comparable to the experimental pump profile (see Sec. \ref{sec:ExpRes}), which can be characterized by a normalized euclidean distance between the radial components of $\|P-P_{\mathrm{exp}}\|/(\|P\|~\|P_{\mathrm{exp}}\|)^{1/2}\approx 0.46$.
    For the simulations, we additionally fix the following system parameters: $m_{\mathrm{eff}}=10^{-4} m_e$ ($m_e$ is the electron mass), $\gamma_c=0.2~\mathrm{ps}^{-1}$, $\gamma_r=1.5 \gamma_c$, $R=0.015~\mathrm{ps}^{-1}~\mu\mathrm{m}^2$, $g_c=6\times 10^{-3}~\mathrm{meV}~\mathrm{\mu m}^2$, and $g_r=2 g_c$.
    Then, the threshold pump intensity for the corresponding homogeneous system is given by $P_{\mathrm{thr}}=\gamma_c\gamma_r/R=4~\mathrm{ps}^{-1}~\mu\mathrm{m}^{-2}$.
    This choice of parameters has been established in our previous work to describe the same semiconductor microcavity sample \cite{MBADNSHS20}, which is introduced in more detail in Sec. \ref{sec:setup}.
    Key simulation results based on these specifications are discussed in the following.

\begin{figure}[t]
	\centering
 	\includegraphics[width=\columnwidth]{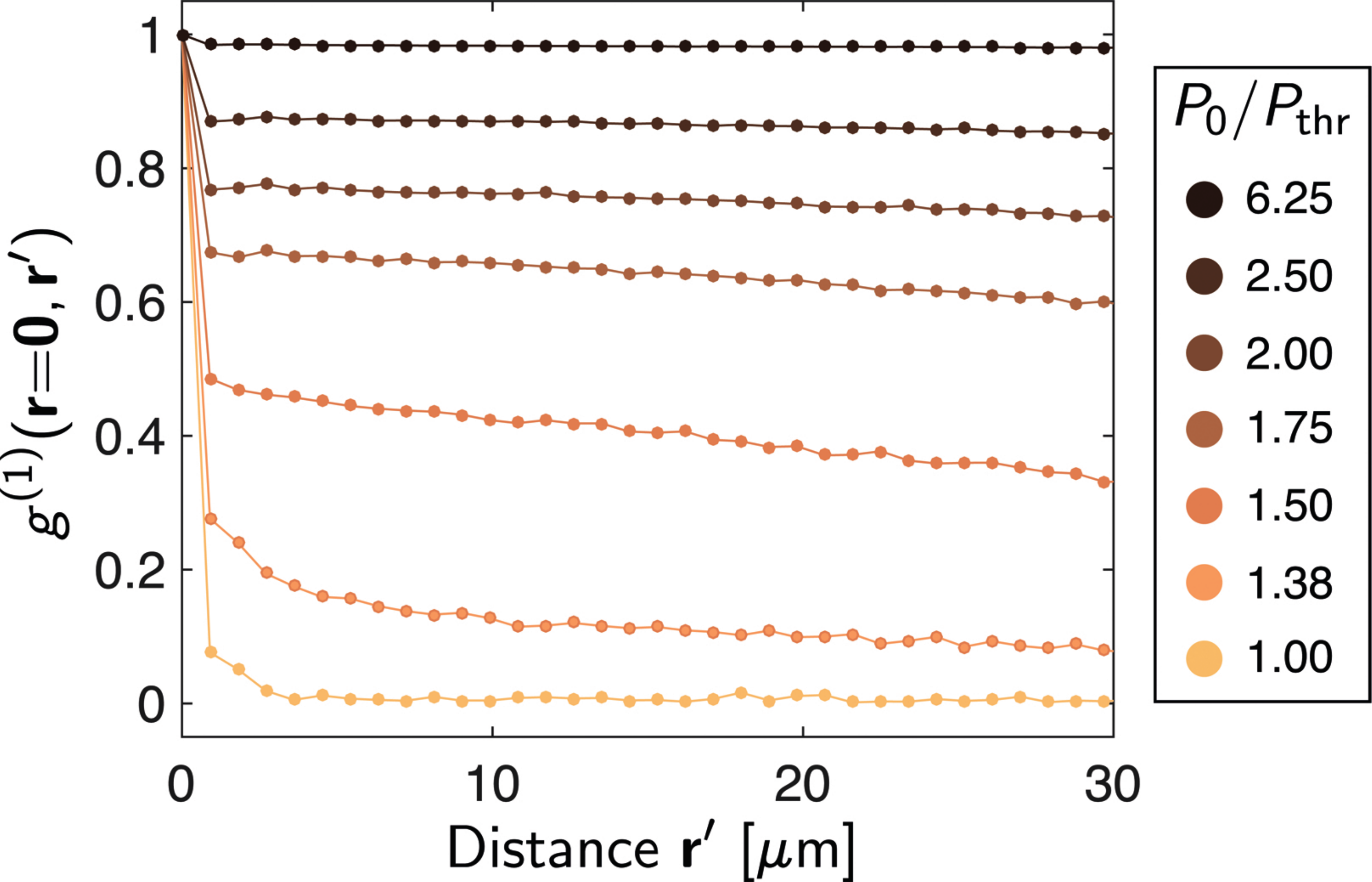}
 	\caption{%
        Simulated first-order, equal-time spatial coherence as a function of the distance from the center of the excitation spot for increasing pump intensities, evidencing the creation of phase coherence across the excitation spot.
    }\label{fig:sim1}
\end{figure}	

    Figure \ref{fig:sim1} shows the simulated first-order equal-time correlation as a function of the distance from the center of the excitation spot,
    \begin{align}
    \begin{aligned}
        g^{(1)}(\mathbf{r},\mathbf{r'})=&\frac{\langle \hat{\psi}^{\dagger}(\mathbf{r})\hat{\psi}(\mathbf{r'}) \rangle}{\sqrt{\langle\hat{\psi}^{\dagger}(\mathbf{r})\hat{\psi}(\mathbf{r})\rangle\langle\hat{\psi}^{\dagger}(\mathbf{r'})\hat{\psi}(\mathbf{r'})\rangle}}.
    \end{aligned}
    \end{align}
    With increasing pump power, the depicted correlation function $g^{(1)}$ increases, implying a buildup of phase coherence across the excitation spot.
    In a broader picture, this kind of coherence presents the basis for all interference phenomena in both the classical and quantum domain \cite{A06,BDW17}.

    At this point, the numerical results evidence the buildup of spatial phase coherence in the polariton system, whose amount clearly depends on the pump power, Fig. \ref{fig:sim1}.
	However, its origin---be it classical or quantum---is not clear at all.
	Importantly, $g^{(1)}$ is not a quantitative measure.
	It certainly provides a length scale over which phase correlations are preserved. 
	But it cannot yield any additional information about the quantum nature, or lack thereof, of the state.
	Quantum coherence as studied in the following, on the other hand, is a direct quantitative measure of the state's resourcefulness from a quantum-computation perspective \cite{SAP17,CG19}.

\begin{figure}[t]
	\centering
 	\includegraphics[width=\columnwidth]{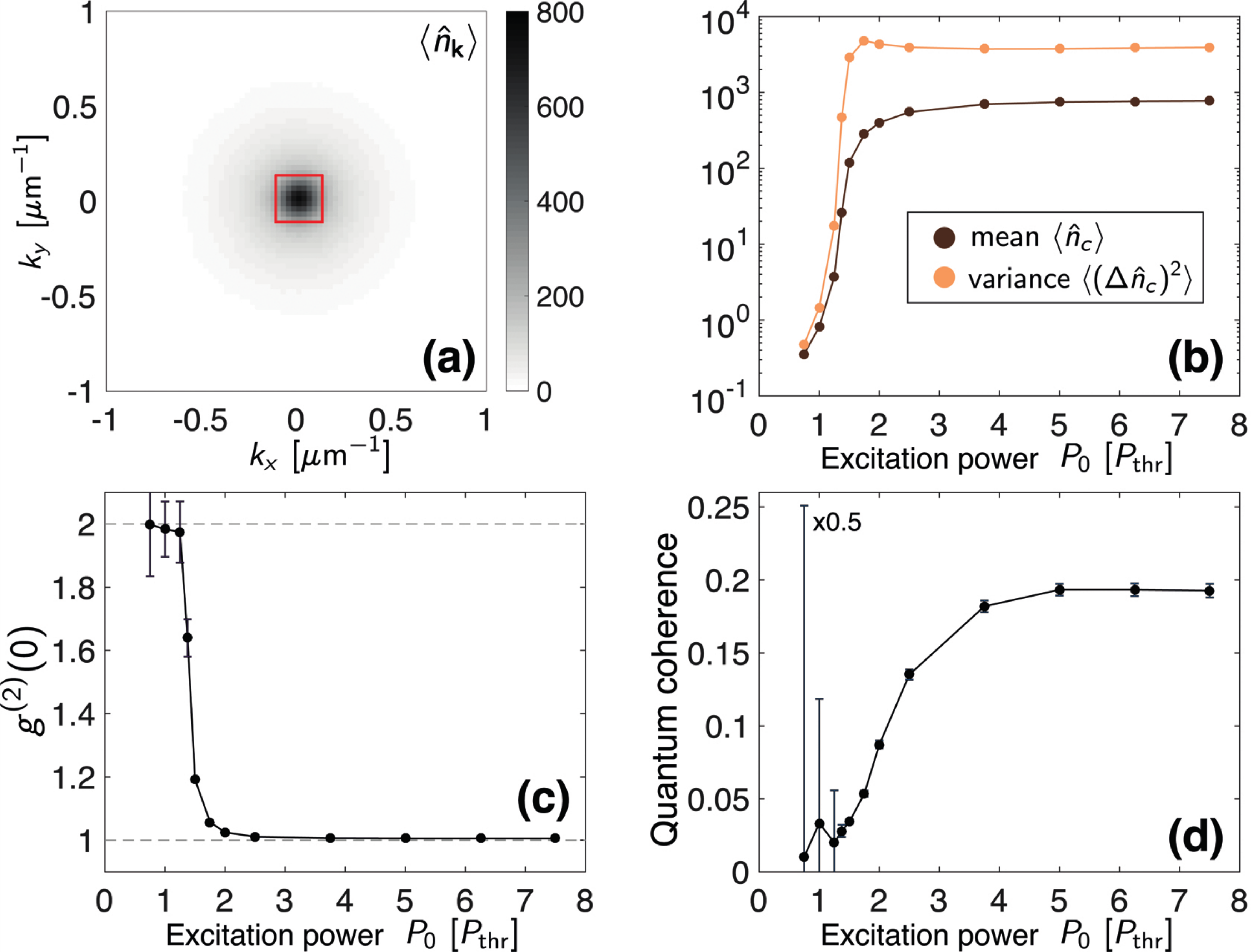}
 	\caption{%
 	    Results from numerical simulations.
        (a) Single-time snapshot of the $k$ space density expectation values above the condensation threshold, $P=2~P_{\mathrm{thr}}$.
        The red square indicates the selected signal area for the mode occupation average. 
        (b) Mean and (c) variance of the averaged polariton excitation number as a function of the pump intensity.
        (c) Equal-time, second-order correlation function $g^{(2)}(\tau=0)$ as a function of the pump intensity, showing the transition from a thermal state to a coherent state across the condensation threshold. 
        (d) Resulting amount of quantum coherence $\mathcal{C}$.
    }\label{fig:sim2}
\end{figure}

    Taking the finite $k$ space linewidth of the condensate emission into account, we average the polariton condensate excitation number across a small square containing $N_p$ discrete modes in the vicinity of $k\approx 0$ [marked in red in Fig. \ref{fig:sim2}(a)], resulting in the following effective number operators and expectation values:
    \begin{subequations}
    \begin{eqnarray}
        \hat{N} &=& \sum_{j=1}^{N_p} \hat{b}^{\dagger}_j\hat{b}_j,
        \\
        \label{eq:n}
        \langle\hat{n}_c\rangle &=& \frac{\langle \hat{N}\rangle}{N_p},
        \\
        \label{eq:var_n}
        \langle (\Delta \hat{n}_c)^2\rangle &=& \frac{\langle (\Delta\hat{N})^2\rangle}{N_p},
    \end{eqnarray}
    \end{subequations}
    where $\hat{b}_j\equiv\hat{b}_{\mathbf{k}_j}$ with $\hat{b}_{\mathbf{k}}=V^{-1/2}\Delta V\sum_{\mathbf{r}}e^{-i\mathbf{k}\mathbf{r}}\hat{\psi}(\mathbf{r})$.

    Figure \ref{fig:sim2}(b) shows the computed mean polariton excitation number and its variance for different excitation powers.
    After a rapid increase above the threshold, both mean and variance saturate for powers $P_0\gtrsim 4~P_{\mathrm{thr}}$.
    Thereby, we predict the transition from a thermal state, $g^{(2)}=2$, to a coherent state, $g^{(2)}= 1$, when increasing the excitation power.
    In this context, a recent photon-number-resolved measurement \cite{KSF18} across the polariton condensation threshold showed a transition from a thermal (i.e., geometric) photon-number distribution to a Poisson distribution of the diagonal elements $p_n$ of the density operator, yet providing no insight into off-diagonal contributions. 
    Applying the model of the displaced thermal state from Sec. \ref{sec:QCoherence} to the polariton-number occupation allows us to calculate coherent ($|\alpha_0|^2$) and thermal ($\bar{n}$) contributions, using the relations \cite{KSF18}
    \begin{subequations}
    \begin{eqnarray}
        \langle\hat{n}_c\rangle &=& \bar{n}+|\alpha_0|^2
        \\
        \text{and}\quad
        \label{eq:var_c}
        \langle(\Delta\hat{n}_c)^2\rangle &=& |\alpha_0|^2(2\bar{n}+1)+\bar{n}^2+\bar{n}.
    \end{eqnarray}
    \end{subequations}
    From those quantities, the value of the second-order, equal-time correlation function $g^{(2)}=g^{(2)}(\tau=0)$ can be determined,
    \begin{equation}
        \begin{aligned}
            g^{(2)}
            =\frac{\langle \hat n_c^2\rangle-\langle \hat n_c\rangle}{\langle \hat n_c\rangle^2}
            % =1+\frac{\langle (\Delta \hat{n}_c)^2\rangle-\langle\hat{n}_c\rangle}{\langle\hat{n}_c\rangle^2} \\
            % =&1+\frac{2|\alpha_0|^2\bar{n}+\bar{n}^2}{(|\alpha_0|^2+\bar{n})^2} \\
            =2-\left(1+\frac{\bar{n}}{|\alpha_0|^2}\right)^{-2},
        \end{aligned}
    \end{equation}
    which was discussed above [see also Fig. \ref{fig:sim2}(c)] and which is a function of the ratio $\bar n/|\alpha_0|^2$ for displaced thermal states.

    Finally, we can determine the measure of quantum coherence from Eq. \eqref{eq:Coherence}.
    Specifically, Fig. \ref{fig:sim2}(d) shows that the coherence increases significantly across the threshold and saturates in a plateau at around
    \begin{equation}
        \label{eq:NumCoherence}
        \mathcal{C}\approx 0.2.
    \end{equation}
    While macroscopic coherence, i.e., $g^{(1)}$, was analyzed before, we here find a nonvanishing amount of quantum coherence that genuinely results from quantum superpositions of particles. 
    This also sets it apart from, for example, $g^{(2)}$ that focuses on very different aspects that pertain to classical waves and nonclassical quantized light.
    But the quantity $g^{(2)}$ fails to provide any insights into the quantum nature of the matter system in which particles are deemed classical.
    In direct opposition, $\mathcal C\gneqq 0$ directly quantifies the amount of quantum superpositions of particles, as described in Sec. \ref{sec:QCoherence}, by assessing the off-diagonal contributions in the density operator in Fock expansion.
	A careful look at Figs. \ref{fig:sim2}(c) and \ref{fig:sim2}(d) further reveals that $g^{(2)}$ and the quantum coherence $\mathcal C$ describe distinct properties.
	For example, the second-order correlation function becomes almost identical to one already at twice the threshold pump power;
	by contrast, one needs approximately 4 times the threshold pump power to reach the maximum quantum coherence in Eq. \eqref{eq:NumCoherence}.

	The saturation value of the quantum coherence depends on the specific system parameters.
	Generally, it is bounded between $0$ and $1$ and increases when the ratio of coherent and thermal contributions increases (see Fig. \ref{fig:QuantCohDTS}).
	The general trend in the simulations is that the quantum coherence slightly increases for increasing condensation rates $R$ and decreasing interaction strengths $g_c$, but it is not very sensitive to those changes.
	Investigations on how to enhance the quantum coherence in polariton condensates, e.g., by using different pump shapes to spatially separate condensate and reservoir or to stabilize orbital-angular-momentum modes, are certainly worth pursuing in future works to optimize the gain in quantum coherence.

%%%%%%%%%%%%%%%%%%%%%%%%%%%%%%%%%%%%%%%%%%%%%%%%%%%%
%%%%%%%%%%%%%%%%%%%%%%%%%%%%%%%%%%%%%%%%%%%%%%%%%%%%
%%%%%%%%%%%%%%%%%%%%%%%%%%%%%%%%%%%%%%%%%%%%%%%%%%%%
\section{Experimental implementation}\label{sec:ExpRes}

\begin{figure}[t]
    \centering
    \includegraphics[width=0.45\textwidth]{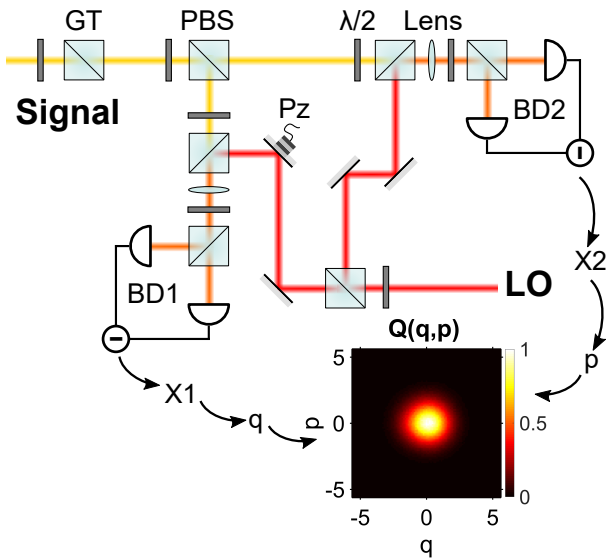}
    \caption{%
        Experimental setup.
        The signal that is emitted by the polariton microcavity is split into two channels.
        Each channel is interfered with the local oscillator (LO), before being detected by a balanced detector (BD).
        The LO path of channel 1 contains a piezo mirror (Pz) for sweeping the relative phase.
        The two BDs yield measurements of the quadratures $X_1$ and $X_2$ that are postselected for retrieving orthogonal quadratures $q$ and $p$, and that are then binned to create a histogram $Q(q,p)$ which corresponds to the Husimi function.
        (PBS, polarizing beam splitter; GT, Glan-Thompson prism; $\lambda/2$, half-wave plate.)
    }\label{fig:setup}
\end{figure}

    In addition to our theoretical investigations, a main feature of this multidisciplinary work is an experimental realization that implements the hybrid light-matter system under consideration.
    Thus, in this section, we detail our setup (Fig. \ref{fig:setup}) and the state reconstruction approach.

%%%%%%%%%%%%%%%%%%%%%%%%%%%%%%%%%%%%%%%%%%%%%%%%%%%%
\subsection{Setup description} \label{sec:setup}

    As a hybrid light-matter system, we study polaritons that are created in a semiconductor microcavity.
    The microcavity features two distributed-Bragg reflectors made of alternating layers of Al$_{0.2}$Ga$_{0.8}$As and AlAs, providing a high quality factor of about $20\,000$.
    These mirrors enclose a $\lambda/2$ cavity, containing four GaAs quantum wells.
    The Rabi splitting of the sample is $9.5~\mathrm{meV}$ and the exciton-cavity detuning is $-6.4~\mathrm{meV}$ for all experiments shown in this work.
    A detailed characterization of the sample can be found in Ref. \cite{MBADNSHS20}.

    The sample is held in a cryostat at $10~\mathrm{K}$ and is excited nonresonantly at the first minimum of the stop band with a continuous-wave laser.
    Note that the nonresonant excitation prevents the system from inheriting coherence from the pump laser since this coherence is completely erased during the energy relaxation process of the polaritons \cite{KRKS02,KMSL07}.
    The spatial profile of the pump laser beam corresponds to a Gaussian spot with a selected diameter of $70~\mathrm{\mu m}$.
    This spot size is chosen to favor condensation at $k = 0$ \cite{KRKS02,WCC08}.
    The emission is spatially filtered by a pinhole with $100~\mathrm{\mu m}$ diameter, corresponding to $13~\mathrm{\mu m}$ on the sample, to ensure the homogeneity of the emission in the investigated region.
    This also corresponds to a filtering in $k$ space around $k = 0$ with a full width at half maximum of $1~\mathrm{\mu m}^{-1}$.

    The properties of the sample emission are investigated via balanced homodyne detection \cite{AR97,RC13}.
    Specifically, we use an eight-port homodyne setup \cite{QLW20} in which the emission is split into two beams.
    Each of them overlaps with a local oscillator (LO) and is sent to a balanced homodyne detector, as shown in Fig. \ref{fig:setup}.
    The polarization of the emission can be chosen with a half-wave plate and a Glan-Thompson prism. 
    The LO, derived from a pulsed Ti:sapphire laser, is resonant with the most intense zero-momentum ground-state mode of the polariton emission.
    Also, the LO shows a Gaussian spatial mode and a full width at half maximum in $k$ space of $1.3~\mathrm{\mu m}^{-1}$, centered at $k = 0$.
    Therefore, the LO overlaps only with signal components around $k \approx 0$.
    Further technical details regarding our unique homodyne detection system can be also found in Ref. \cite{LTA18}.

    Field quadratures of the emission $X_1$ and $X_2$ are measured in two channels through the balanced detectors BD1 and BD2 in Fig. \ref{fig:setup}. 
    The relative phase between the two channels is swept continuously by a piezo mirror in one of the LO paths.
    In postprocessing, we remove spurious correlations between consecutive quadrature measurements \cite{KBMCHL12} and normalize the result with respect to the LO amplitude for obeying the commutator convention $[\hat q,\hat p]=i$.
    Then, we keep only data where the product $X_{1}X_{2}$ equals zero within a $\pm2.5\%$ margin of the peak-to-peak value.
    In this case, the quadratures are orthogonal and hence represent the phase-space coordinates $q$ and $p$, resembling the real and imaginary part of the coherent amplitude $\alpha$ up to a  constant factor.

    In a next step, we assemble a histogram of the occurrences of pairs $(q,p)$, resulting in the Husimi $Q$ distribution \cite{S01}.
    As the phase of the sample light field fluctuates on a timescale of $100~\mathrm{ps}$, given by its coherence time, there is no long-term phase stability between the signal and the LO.
    Therefore, in a measurement that takes several hundred milliseconds to accumulate enough quadrature values, we average over all phases between signal and LO and receive a phase-averaged version of the actual $Q$ function.
    The phase information could be obtained by employing more homodyne channels and reconstructing the relative phase between them \cite{TLA20}.
    However, the phase-averaged $Q$ function already provides the necessary information since Eq. \eqref{eq:Coherence} requires only $\bar n$ and $|\alpha_0|^2$, but not the phase $\arg\alpha_0$.

%%%%%%%%%%%%%%%%%%%%%%%%%%%%%%%%%%%%%%%%%%%%%%%%%%%%
\subsection{Phase-space reconstruction}

\begin{figure}[b]
    \centering
    \includegraphics[width=0.5\textwidth]{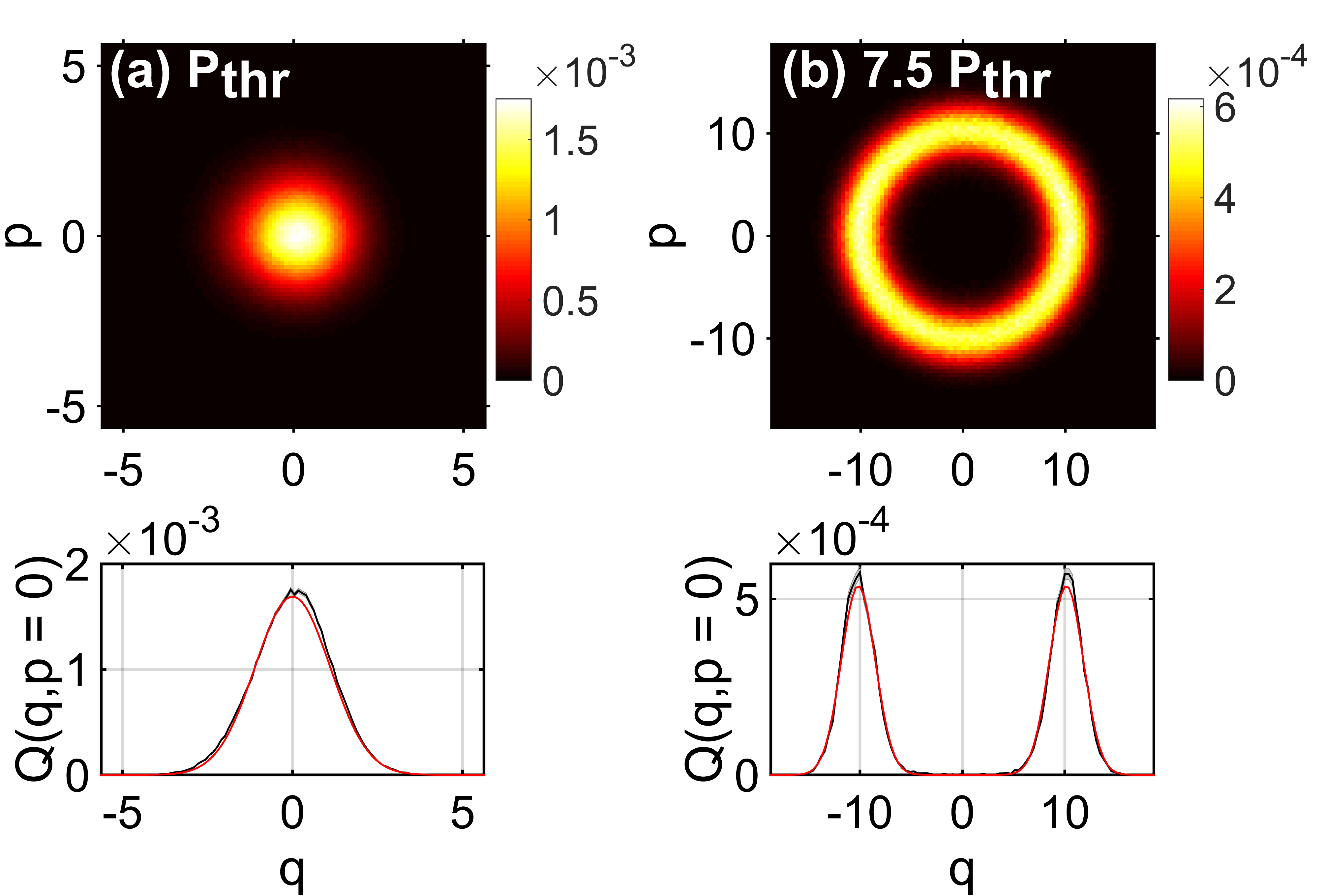}
    \caption{%
        Reconstructed phase-averaged Husimi function (a) for $P = P_\mathrm{thr}$ and (b) for $P = 7.5~P_\mathrm{thr}$.
        The bottom plots depict cuts along the $q$ axis ($p = 0$).
        Black lines represent data while red curves show the fitted function of a displaced thermal state.
        A one standard-deviation error margin, given by the standard error of the counting statistics, is shown as shaded areas but is mostly not visible.
    }\label{fig:Husimis}
\end{figure}

    An example of two reconstructed phase-averaged Husimi functions for different excitation powers is displayed in Fig. \ref{fig:Husimis}.
    For these measurements, the linear polarization has been selected to be parallel to one of the crystal axes (henceforth called vertical polarization).
    The top panel of Fig. \ref{fig:Husimis}(a) shows the $Q$ function at the threshold power $P_\mathrm{thr} = 30~\mathrm{mW}$.
    Under these conditions, the phase-averaged Husimi function is, in fact, a Gaussian distribution centered at the phase-space origin, as shown by the fit.
    As already stated in Sec. \ref{sec:setup}, both orthogonal quadratures, $p$ and $q$, are directly related to the coherent amplitude $\alpha$ and the number of coherent photons of the measured light field.
    Therefore, a phase distribution centered at $q=0$ and $p = 0$ indicates an almost inexistent coherent population $|\alpha_0|^2 \approx 0$ and can be identified as a signature of an incoherent thermal state.

    In contrast, for a higher power of $7.5~P_\mathrm{thr}$ [Fig. \ref{fig:Husimis}(b), top panel], the phase-space function clearly shows a displacement, resembling a ring with a radius of $(q_{0}^{2}+p_{0}^{2})^{1/2} = 10.4$ in quadrature units when phase averaged.
    This displacement corresponds to the coherent photon number $|\alpha_0|^2 = (q_{0}^{2}+p_{0}^{2})/2 = 53$ whereas the ring width stems from the thermal contribution, $\bar{n} = 1.7$.
    Thus, the measured $Q$ functions match the expected phase-averaged displaced thermal state, which can be quantified by fitting them to the model described in Eq. \eqref{eq:HusimiPDTS} in Appendix \ref{app:DisThState}.
    This fit is depicted in the lower panels of Fig. \ref{fig:Husimis} (red line) alongside with the data (black line) for a cut along the $q$ axis ($p = 0$).
    It can be seen that the fit matches our data well.
    The comparison of the functions at both excitation powers allows us to observe a clear phase transition from a noncondensed, thermal state to a condensed, coherent state.
    The appearance of this spontaneous coherence in polaritonic systems has been studied and identified as a signature of Bose-Einstein condensation \cite{DWSBY02,KRKS02,BHSPW07}.

    The phase-averaged Husimi distributions shown in Fig. \ref{fig:Husimis} have been obtained by binning all quadratures that we recorded over several hundred milliseconds.
    This only works if the state of the system remains sufficiently constant over time.
    However, semiconductor systems often show nontrivial dynamics, resulting in nonconstant states of their emission.
    This is the case in our system for a small range of intermediate powers, i.e., from $5.6~ P_\mathrm{thr}$ up to $6.8~P_\mathrm{thr}$. 
    In this power range, the emission alternates in time between a state with high intensity and one with low intensity.

\begin{figure}[b]
    \centering
    \includegraphics[width=0.5\textwidth]{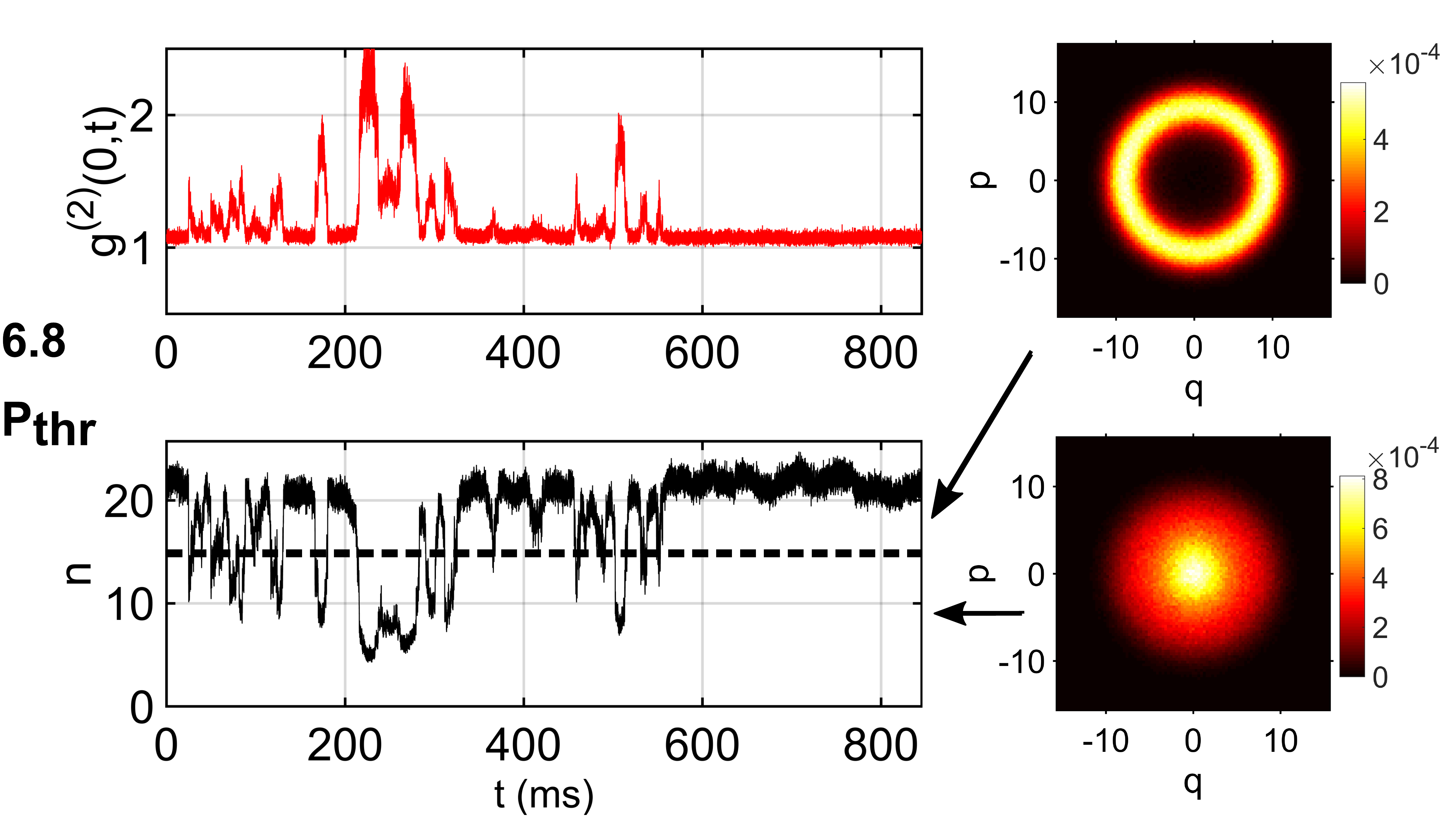}
    \caption{%
        Time-resolved photon number and $g^{(2)}(\tau = 0,t)$ for $P = 6.8~P_\mathrm{thr}$.
        The initial time $t = 0$ is arbitrarily chosen as the beginning of the measurement. 
        A dashed line in the bottom panel marks the frontier between the state with high intensity and the one with low intensity.
        Right panels show the phase-averaged Husimi functions above (top plot) and below (bottom plot) this limit.
    }\label{fig:switching}
\end{figure}

    This bistability is revealed by time-resolved measurements of the mean photon number and $g^{(2)}(\tau = 0,t)$ \cite{LTA18} in Fig. \ref{fig:switching}.
    In the high state, $g^{(2)}(\tau = 0)$ is close to 1, indicating mostly coherent emission.
    In the low state, the value approaches two, corresponding to a high amount of thermal emission.
    Thus, an averaging over all data available for these pump powers to produce one single Husimi function does not adequately represent the state of the system.
    Rather, it is sensible to consider the Husimi functions for instances with high and low intensity separately.
    To this end, we define ``high'' as the state when the momentary photon number is larger than the mean of the maximum and minimum photon number of the data set whereas ``low'' is defined as the state when the momentary photon number is smaller than that.
    The boundary between both states is marked with a dashed line in the bottom panel of Fig. \ref{fig:switching}.
    As examples, the images of the phase-averaged Husimi functions below and above this limit are also shown.
    Therefore, in this range of intermediate powers, we perform a fit for each measurement and obtain different sets of parameters for the high and low states.
    Thus, our time-resolved detection method reveals an interesting and nontrivial bistability behavior of the semiconductor dynamics that we are able to adequately account for with our experimental analysis.

%%%%%%%%%%%%%%%%%%%%%%%%%%%%%%%%%%%%%%%%%%%%%%%%%%%%
\section{Results and discussion}\label{sec:ResultsDiscuss}

\begin{figure*}
    \centering
    \includegraphics[width=0.9\textwidth]{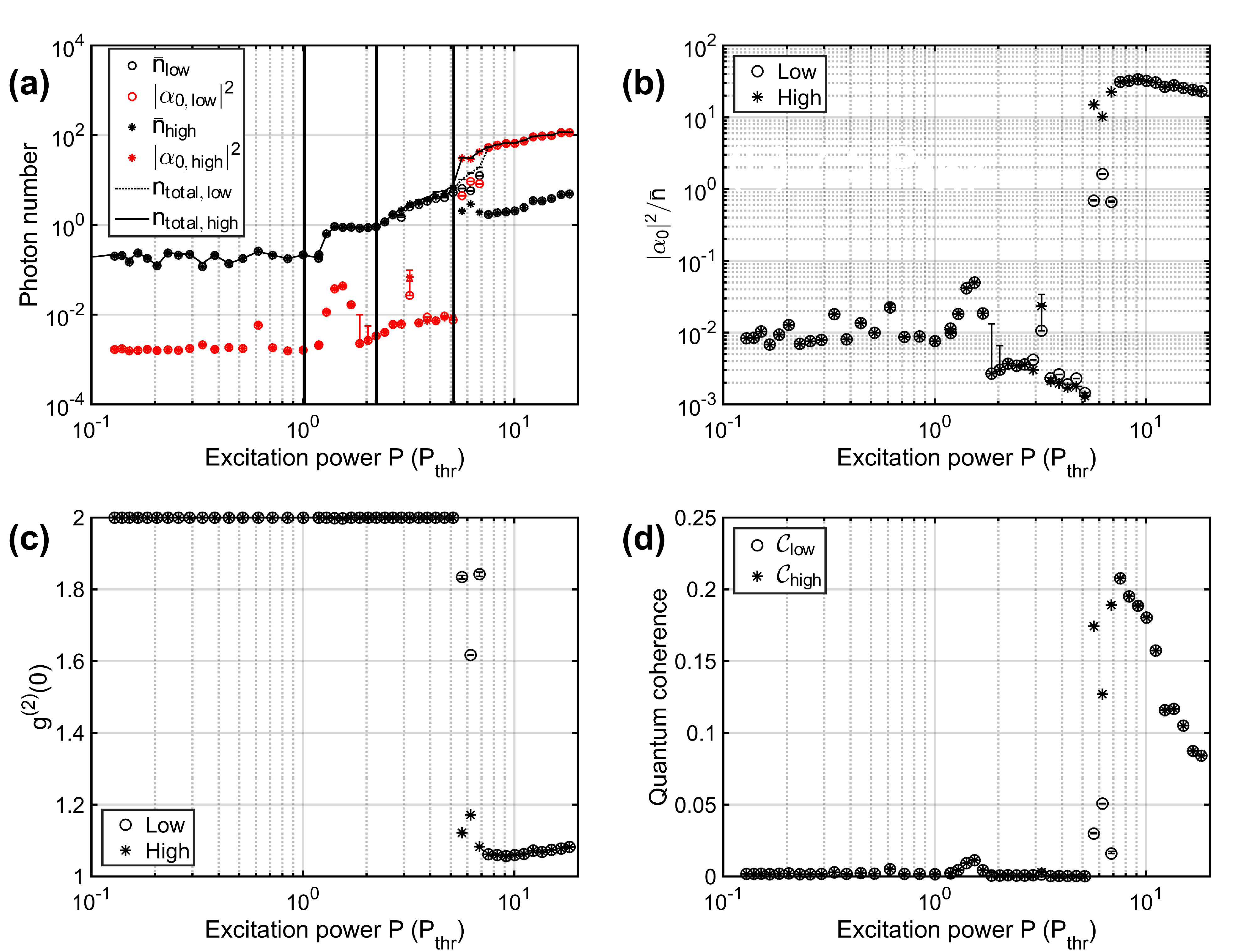}
    \caption{%
        Results derived from fitting phase-averaged Husimi functions as a function of the pump intensity for vertical polarization.
        (a) Coherent (red circles) and thermal (black circles) photon numbers.
        Black line shows the total photon number $n_{\mathrm{total}}$. 
        Thick, black vertical lines indicate thresholds between different regimes of emission.
        (b) Ratio between the coherent and thermal photon-number contribution.
        (c) Equal-time, second-order correlation function $g^{(2)}(\tau=0)$.
        (d) Amount of quantum coherence.
        Open and closed symbols correspond to the low and high state, respectively, when the emission is switching on and off.
        When both symbols overlap, the emission is stable.
        Error bars are obtained from a Monte Carlo error propagation (see Appendix \ref{app:NumDetails} for details);
        because of the asymmetry of the logarithmic scale, only the upper part of the error margin is depicted.
    }\label{fig:2020-10-06-n-g2-Coherence-l2-0}
\end{figure*}

\begin{figure*}
    \centering
    \includegraphics[width=0.7\textwidth]{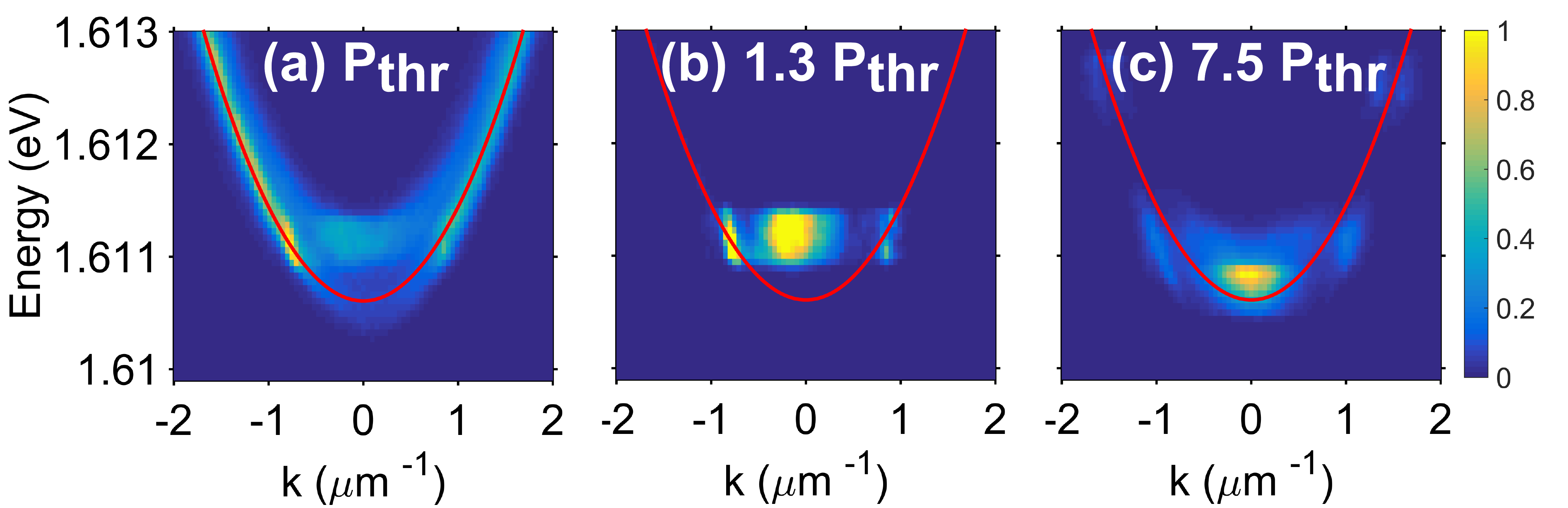}
    \caption{% 
        Dispersion curves measured at different excitation powers.
        In each panel, the emission is normalized to one and plotted on a linear color scale.
        The red line shows the lower polariton dispersion obtained from fitting the dispersion at a power below threshold.
        In plot (a) at $P_\mathrm{thr}$ and in (b) at $1.3~P_\mathrm{thr}$, the emission around $k = 0$ is blueshifted because of interaction with the reservoir.
        In (b), emission at $k = \pm 0.8~\mathrm{\mu m}^{-1}$ is visible.
        In panel (c), $7.5~P_\mathrm{thr}$, this side emission is mostly suppressed when compared with the dominant emission at $k = 0$, which is redshifted because of heating.
    }\label{fig:dispersions}
\end{figure*}

    The aforementioned fits of the phase-averaged Husimi functions (Fig. \ref{fig:Husimis}) yield the thermal photon number $\bar n$ and the coherent photon number $|\alpha_0|^2$ of the measured light field.
    In addition, the total photon number $n_{\mathrm{total}}$ can be obtained.
    These photon numbers are plotted as a function of the excitation power in Fig. \ref{fig:2020-10-06-n-g2-Coherence-l2-0}(a) for vertical polarization.
    The ratio of coherent and thermal photon number is additionally shown in Fig. \ref{fig:2020-10-06-n-g2-Coherence-l2-0}(b).
    From the photon numbers, we can derive the variance $(\Delta n_{\mathrm{total}})^2$ via Eq. \eqref{eq:var_c} and  $g^{(2)}(\tau = 0)$ from that [Fig. \ref{fig:2020-10-06-n-g2-Coherence-l2-0}(c)].
    Importantly, the amount of quantum coherence $\mathcal{C}$, Eq. \eqref{eq:Coherence}, can be estimated as well, Fig. \ref{fig:2020-10-06-n-g2-Coherence-l2-0}(d).
    Our data further enable us to analyze how the plotted quantities depend on the pump power to describe the transition from an incoherent ensemble of Fock states to states with a significant amount of quantum coherence.

    For excitation powers below the threshold, the emission is almost completely thermal.
    This is evidenced by a difference of 2 orders of magnitude between $\bar n$ and $|\alpha_0|^2$ and a correlation function value $g^{(2)}(\tau = 0) = 2$.
    Importantly, the determined amount of quantum coherence is close to zero.
    This evidences an incoherent state of polariton particles in the sample, which have not yet reached the critical density for condensation.
    The thermal phonon bath induces polariton relaxation \cite{HSQHFY10, SQ08}, and the polaritons cannot directly inherit the coherence from a nonresonant excitation.
    Therefore, they remain in an incoherent state without quantum superpositions.

    At the threshold power, both the thermal and coherent photon numbers rise.
    In particular, the coherent photon number $|\alpha_0|^2$ shows a significant increase, implying a corresponding increase of coherence in the semiconductor.
    The number of polaritons in the ground state approaches 1, indicating the onset of quantum degeneracy and stimulated scattering to the ground state, which initializes the condensation process.
    Notably, the quantum coherence $\mathcal{C}$ slightly increases as well but is also affected by the increase of the thermal background.

    With further increasing power, the total photon number of the measured light field stays on a plateau until $2~P_\mathrm{thr}$ is reached.
    In this regime, the coherent photon number drops again and coherence is lost.
    The thermal photon number is much higher, leading to $g^{(2)}(\tau = 0) = 2$.
    A possible reason for this is mode competition of the ground state with modes at higher $k$ values, which do not overlap with the LO and are thus not measured with our homodyne detection setup.
    It is worth emphasizing that this mode selectivity provided by the LO is desirable in order to specify which mode of the system we are investigating with regard to quantum coherence.
    Additional modes are revealed in the dispersions shown in Figs. \ref{fig:dispersions}(a) and \ref{fig:dispersions}(b), where two new contributions to the emission appear at $k = \pm 0.8~\mathrm{\mu m}^{-1}$ for powers close to the threshold.
    We remark that these are not covered by the LO and additionally filtered out by the pinhole in the signal beam.
    This emission originates from polaritons that are repelled by the excitonic reservoir and scattered to higher $k$ states, preventing them from being trapped at $k = 0$.
    Particularly for negative detuning, energy relaxation of polaritons is less efficient, leading to the accumulation of polaritons at high-$k$ states under continuous-wave excitation \cite{EGBO18}.
    Thus, the critical density is not reached, and a condensate at $k = 0$ cannot form yet.

    For a power that exceeds $2~P_\mathrm{thr}$, the emission from higher-$k$ states fades away and the emission from $k = 0$ predominates, leading to an increase in $n_{\mathrm{total}}$.
    Still, the coherence remains on a comparably low level.
    Here, the vertical polarization competes with the horizontal one.
    Only for powers between $5~P_\mathrm{thr}$ and $7~P_\mathrm{thr}$, the coherent photon number exceeds the thermal photon number, $g^{(2)}(\tau = 0)$ drops towards 1, and the quantum coherence rises significantly.
    Hence, the density of polaritons at $k = 0$ becomes sufficient for condensation.
    In this power range, however, the emission switches between a state with high intensity and one with low intensity, as shown in Fig. \ref{fig:switching}.
    This bistability occurs due to the mode competition with the horizontal polarization mode, being caused by sample disorder \cite{KRKS02}.

    Eventually, for powers well above $7.5~P_\mathrm{thr}$, the vertical polarization mode wins the gain competition, and a stable coherent emission at $k = 0$ is reached.
    This can be additionally seen in the dispersion in Fig. \ref{fig:dispersions}(c).
    Here, the coherence also takes its maximum value, $\mathcal{C} = 0.208\pm 0.001$; see Fig. \ref{fig:2020-10-06-n-g2-Coherence-l2-0}(d).
    Consequently, the system now carries a highly significant amount of quantum coherence.
    In fact, this value fits closely to the one expected from our simulation, $\mathcal{C}=0.2$;
    see Sec. \ref{sec:NumSim} and Eq. \eqref{eq:NumCoherence}.
    In contrast to nonclassicality in the context of quantum optics that is commonly adapted for condensed-matter systems, our form of quantum coherence tells us that there are highly non-negligible quantum superpositions of polariton particles in the condensate, which are a result of off-diagonal density operator components and quantified by $\mathcal C>0$.
    Recall that the family of displaced thermal states does not carry any nonclassicality (because of $P\geq0$) in the sense for quantized radiation fields.
    Still, quantum coherence of the polaritons is successfully probed by identifying superpositions of Fock states in the emitted light, as elaborated in Sec. \ref{sec:QCoherence}.

    Finding quantum coherence in the polariton condensate is significant as it, for example, renders a conversion into entanglement and other quantum correlations possible \cite{KSP16,CH16,Qetal18,MYGVG16}.
    As such, the measured quantum coherence is an operationally equivalent quantum-computational resource as entanglement \cite{SAP17,WY16,CG19}, ans is thus highly sought after for quantum-information protocols.
    That is, the superpositions of Fock states can be transformed into entanglement without requiring coherent processes where the amount of output entanglement is identical to the input coherence \cite{KSP16}.
    For example, a parametric pair-production process, $|n\rangle\mapsto|n\rangle\otimes|n\rangle$, can be used for this purpose; see also Appendix \ref{app:EntanglementMonotonocity} for technical details.
    Similar parametric pair scattering processes in polariton systems include the polariton optical parametric oscillator \cite{BSS00,ASLB09,SCT12} and quantum depletion \cite{PEB20}.
    It is also worth emphasizing that quantum coherence exclusively quantifies superpositions that stem from the linear combination of quantum-state vectors, here number states, as discussed in Sec. \ref{sec:QCoherence}.
    This is in contrast to other forms of coherence, which can originate from both the classical interference of waves and quantum superpositions alike.

    Therefore, our findings demonstrate the general capabilities of the studied system to provide states that are resourceful enough for quantum-processing tasks beyond classical coherence.
    It is, however, important to remark that we explicitly do not claim that we carried out a quantum protocol.
    Rather, the notion of quantum coherence is applied as a means to quantify the general usefulness of the generated states in such protocols within the well-established framework of quantum-resource theories.
    Furthermore, the main result here is the demonstration of the interface aspect of the light-matter system by successfully probing particle quantum superpositions in the condensate via the emitted quantum light.

    In general, the higher the coherence, where $0\leq\mathcal C\leq1$, the higher the state's value as a quantum resource.
    Clearly, the generated coherence does not saturate the upper boundary of one in our conceptual demonstration.
    Nevertheless, the reached value $\mathcal{C} = 0.208\pm 0.001$ is highly significant and, thus, shows a clear deviation from the classical lower bound zero, implying a non-negligible resourcefulness of the produced state in quantum-information processing.
    Future optimizations as mentioned in Sec. \ref{sec:NumSim} are planned to further increase the accessible amount of quantum coherence.

    Eventually, when further increasing the pump power, the coherent photon number approaches saturation while the thermal photon number still grows.
    Consequently, $g^{(2)}(\tau = 0)$ increases slightly and the coherence again drops below 0.1.
    This decrease of coherence for high powers can be explained by polariton-polariton scattering \cite{SQ08} and by the heating of the sample since coupling to the thermal bath of the lattice and carriers causes decoherence.
    The heating of the sample also manifests in a redshift of the polariton emission, which can be clearly observed when comparing (b) and (c) in Fig \ref{fig:dispersions}.

    Finally, we can also briefly compare the numerical simulation of the semiconductor system, Figs. \ref{fig:sim2}(c) and \ref{fig:sim2}(d), with the measurements of the emitted light, resulting in Figs. \ref{fig:2020-10-06-n-g2-Coherence-l2-0}(c) and \ref{fig:2020-10-06-n-g2-Coherence-l2-0}(d).
    In general, both experiment and simulation show a transition from a thermal to a nearly perfect coherent state, indicated by a rapid drop of $g^{(2)}(\tau = 0)$ from 2 to 1 and an increase of the quantum coherence $\mathcal C$.
    While the reduction in $g^{(2)}$ indicates only a transition of the photon statistics from a thermal to a Poisson one, the buildup of coherence addresses the increase of off-diagonal density matrix contributions as a result of quantum superpositions.
    A few differences between experiment and simulation are explained by the effects of mode competition in $k$ space, polarization mode competition, heating, and sample disorder, which are not accounted for in the simulation.
    For instance, due to mode competition, condensation at $k=0$ happens at higher powers in the experiment as compared to the simulation, and due to the heating, the quantum coherence does not saturate in a plateau.
    In addition, we have to keep in mind that the mapping of the polariton properties to the measured light is bijective but not an identity, as shown in Appendix \ref{app:LinCoupling} where the corresponding phase-space distributions are distinguishable by a rescaled argument alone.
    However, both experiment and simulation agree in the central result that is the amount of quantum coherence carried by the hybrid system. 

%%%%%%%%%%%%%%%%%%%%%%%%%%%%%%%%%%%%%%%%%%%%%%%%%%%%
%%%%%%%%%%%%%%%%%%%%%%%%%%%%%%%%%%%%%%%%%%%%%%%%%%%%
%%%%%%%%%%%%%%%%%%%%%%%%%%%%%%%%%%%%%%%%%%%%%%%%%%%%
\section{Conclusion}\label{sec:Conclusion}

    In our multidisciplinary work, encompassing quantum-information science, semiconductor physics, and quantum optics, we theoretically and experimentally study quantum interference phenomena in polariton condensates by phase-space reconstructions of the state of light that carries the information about particle superpositions within the probed condensate.
    In contrast to other nonclassicality criteria used in condensed-matter physics, the notion of quantum coherence applied here considers particle aspects of the system as classical, and superpositions of such Fock states are identified as resourceful states for quantum-information applications.
    On the one hand, our results demonstrate the quantum-coherent interfacing capabilities of two distinct physical platforms, polariton system and emitted light, thereby addressing the challenging problem of interconnecting different physical devices in a manner that is useful for quantum-information technologies.
    On the other hand, we quantify a vital resource for quantum-information processing---quantum coherence---in polariton condensates.
    Akin to the emission characteristics of a laser, we report on the successive buildup of quantum coherence in our system when increasing the power of the classical pump field;
    whilst below threshold, an incoherent thermal behavior of the polaritons is observable.
    Beyond that, however, we further demonstrate that the transition from incoherent to coherent behavior presents a much richer structure in our system, such as bistabilities, than one expects from the much simpler physics of a laser.

    In our matter system, particles represent the classical reference and their superpositions define quantum effects, unlike for electromagnetic waves where bare photons are already considered to be nonclassical.
   	Therefore, commonly applied quantumness criteria for light cannot be applied to witness the quantum features of a condensate.
	Rather, polariton superpositions are encoded in the superposition of photon-number states.
	This very relation is used as a proxy characterization of the semiconductor system, whose quantum coherences are not directly accessible.
	By determining superpositions of photon-number states, we are then able to quantify the quantum nature of the matter system under study.
	To this end, we combine modern continuous-variable phase-space simulation and reconstruction techniques with recently developed, discrete-variable-based methods to quantify quantum coherence in the superposition of quantum particles.
	The predicted and experimentally determined amount of quantum coherence agree with each other.

	In the future, we plan to investigate the discussed effects of the polarization-mode competition in greater detail, in simulation and experiments, as it hints at interesting quantum-physical effects that are not expected for scalar quantum fields.
	In addition, our current experiment is not able to resolve the phase, resulting in phase-averaged Husimi functions of displaced thermal states.
	But our simulation of correlation functions for the polariton system directly implies that this averaging does not describe the true state.  
	Thus, a targeted improvement of our setup is enabling a phase-resolved measurement in the future, which is, for example, also vital for the aforementioned polarization analysis.

    In conclusion, our multidisciplinary research connects different fields and arches across different physical platforms, which is essential for the future application and success of quantum-information science and technology.
    At the same time, we study fundamental superposition effects with the potential for discovering new quantum phenomena in hybrid systems.
    Our proof-of-concept demonstration of a classical-to-quantum transition presents a promising starting point for future research and applications.

%%%%%%%%%%%%%%%%%%%%%%%%%%%%%%%%%%%%%%%%%%%%%%%%%%%%
%%%%%%%%%%%%%%%%%%%%%%%%%%%%%%%%%%%%%%%%%%%%%%%%%%%%
%%%%%%%%%%%%%%%%%%%%%%%%%%%%%%%%%%%%%%%%%%%%%%%%%%%%
\begin{acknowledgments}
    The authors gratefully acknowledge funding through the Deutsche Forschungsgemeinschaft (DFG, German Research Foundation) via the transregional collaborative research center TRR~142 (Project A04, Grant No. 231447078). A grant for computing time at the Paderborn Center for Parallel Computing (PC2) is gratefully acknowledged.
\end{acknowledgments}

%%%%%%%%%%%%%%%%%%%%%%%%%%%%%%%%%%%%%%%%%%%%%%%%%%%%
%%%%%%%%%%%%%%%%%%%%%%%%%%%%%%%%%%%%%%%%%%%%%%%%%%%%
%%%%%%%%%%%%%%%%%%%%%%%%%%%%%%%%%%%%%%%%%%%%%%%%%%%%
\appendix

%%%%%%%%%%%%%%%%%%%%%%%%%%%%%%%%%%%%%%%%%%%%%%%%%%%%
\section{Supplemental details on quantum coherence}\label{app:QuantCoh}

    In this Appendix, we prove several relations that are used in the main text.
    These proofs are straightforward but offered here for a self-consistent reading of this work.

    Firstly, we compute the Hilbert-Schmidt distance of a state, $\hat\rho=\sum_{m,n}\rho_{m,n}|m\rangle\langle n|$, and its incoherent counterpart, $\hat\rho_\mathrm{inc}=\sum_n\rho_{n,n}|n\rangle\langle n|$ \cite{SW18}.
    In doing so, we find the following, simplified formula for the amount of quantum coherence:
    \begin{equation}
    \begin{aligned}
        &\|\hat\rho-\hat\rho_\mathrm{inc}\|_\mathrm{HS}^2
        =\mathrm{tr}\left(\left[\hat\rho-\hat\rho_\mathrm{inc}\right]^2\right)
        \\
        =&\mathrm{tr}\left(\hat\rho^2\right)
        -2\,\underbrace{
            \mathrm{tr}\left(\hat\rho\hat\rho_\mathrm{inc}\right)
        }_{\substack{
            =\sum_n \rho_{n,n}\langle n|\hat\rho|n\rangle
            \\
            =\mathrm{tr}\left(\hat\rho_\mathrm{inc}^2\right)
            \hspace{4.5ex}
        }}
        +\mathrm{tr}\left(\hat\rho_\mathrm{inc}^2\right)
        \\
        =&\mathrm{tr}\left(\hat\rho^2\right)
        -\mathrm{tr}\left(\hat\rho_\mathrm{inc}^2\right).
    \end{aligned}
    \end{equation}

    Secondly, in terms of phase-space representations, we now show that the best incoherent approximation $\hat\rho_\mathrm{inc}=\sum_n\rho_{n,n}|n\rangle\langle n|$ \cite{SW18} is obtained by a phase-averaging over the corresponding phase-space distribution.
    Say we have $\hat\rho=\int d^2\alpha\, P(\alpha)|\alpha\rangle\langle\alpha|$ and $\hat\rho_\mathrm{inc}=\int d^2\alpha\, P_\mathrm{inc}(\alpha)|\alpha\rangle\langle\alpha|$, where $P_\mathrm{inc}(\alpha)=(2\pi)^{-1}\int_0^{2\pi} d\vartheta P(\alpha e^{i\vartheta})$.
    It is known that the resulting phase-insensitive distribution $P_\mathrm{inc}$ is diagonal in the number basis \cite{VW06}.
    Thus, what is left to show is that the diagonal elements of both density matrices are identical.
    For this purpose, we compute
    \begin{equation}
    \begin{aligned}
        &\langle n|\hat\rho|n\rangle
        =\int d^2\alpha\,P(\alpha)|\langle\alpha|n\rangle|^2
        \\
        =&\int_0^{\infty} dr\,r
        \underbrace{
            \int_{0}^{2\pi} d\varphi\, P(re^{i\varphi})
        }_{
            =\int_{0}^{2\pi} d\varphi\, P_\mathrm{inc}(re^{i\varphi})
        }
        \frac{r^{2n}}{n!}e^{-r^2}
        =\langle n|\hat\rho_\mathrm{inc}|n\rangle.
    \end{aligned}
    \end{equation}
    Note that the same relation follows for the Wigner and Husimi functions since they are obtained from the Glauber-Sudarshan distribution via a convolution with a radially symmetric Gaussian kernel that is centered at zero.

    Third, the purities, which are relevant for computing the quantum coherence, can be expressed in terms of phase-space distributions.
    Note that it suffices to restrict to general states $\hat\rho$ as the same then applies to phase-averaged states, i.e., $\hat\rho_\mathrm{inc}$.
    We can write the following equivalent expressions:
    \begin{equation}
    \begin{aligned}
        \mathrm{tr}\left(\hat\rho^2\right)
        =&\int d^2\alpha\,d^2\alpha'\,P(\alpha)P(\alpha')\exp\left(-|\alpha-\alpha'|^2\right)
        \\
        =&\pi\int d^2\alpha P(\alpha)Q(\alpha)
        =\pi\int d^2\alpha \left[W(\alpha)\right]^2,
    \end{aligned}
    \end{equation}
    where we used the overlap $|\langle\alpha|\alpha'\rangle|^2=e^{-|\alpha-\alpha'|^2}$, the representation $Q(\alpha)=\langle\alpha|\hat\rho|\alpha\rangle/\pi$ for the Husimi function, and general Gaussian integral transformations of so-called $s$-parametrized phase-space distributions for the relation that includes the Wigner function $W(\alpha)$ \cite{VW06}.

%%%%%%%%%%%%%%%%%%%%%%%%%%%%%%%%%%%%%%%%%%%%%%%%%%%%
\section{Properties of displaced thermal states}\label{app:DisThState}

    Here, we provide further details on the family of states under study.
    These displaced thermal states are defined through their Glauber-Sudarshan representation as
	\begin{equation}
		\hat\rho=\int d^2\alpha\, \frac{\exp\left(-\frac{|\alpha-\alpha_0|^2}{\bar n}\right)}{\pi\bar n} |\alpha\rangle\langle \alpha|,
	\end{equation}
	where $\alpha_0\in\mathbb C$ and $\bar n>0$.
	Note that the probability density approaches a Dirac delta distribution for $\bar n\to 0^{+}$.
    In addition, the Husimi function can be straightforwardly obtained,
	\begin{equation}
		Q(\alpha)
		=\frac{\langle \alpha|\hat\rho|\alpha\rangle}{\pi}
		=\frac{\exp\left[-\frac{|\alpha-\alpha_0|^2}{\bar n+1}\right]}{\pi(\bar n+1)},
	\end{equation}
	and the Wigner function takes again such a Gaussian form, however, with a variance $\bar n+1/2$.

    For fitting our experimental data, one can readily compute the phase-averaged version of the above distributions, e.g., $P_\mathrm{inc}(\alpha)=[2\pi^2\bar n]^{-1}\int_0^{2\pi}d\varphi\,\exp\left[-\left|\alpha-\alpha_0e^{i\varphi}\right|^2/\bar n\right]$ for the Glauber-Sudarshan distribution.
    For example, in the case of the Husimi function, we find the explicit formula
	\begin{equation}
	\label{eq:HusimiPDTS}
	    Q_\mathrm{inc}(\alpha)=\frac{\exp\left(-\frac{|\alpha|^2+|\alpha_0|^2}{\bar n+1}\right)}{\pi(\bar n+1)} 
	    I_0\left(\frac{2|\alpha||\alpha_0|}{\bar n+1}\right).
	\end{equation}
	Herein, $I_0$ denotes the zeroth modified Bessel function of the first kind, $2\pi\, I_0(\lambda)=\int_0^{2\pi} d\varphi\,e^{\lambda\cos(\varphi)}$.
	We remark that because of the phase independence, one obtains a vanishing mean coherent amplitude, $\langle\hat a\rangle=0$, for phase-averaged states.

    For obtaining correlation functions, we introduce the quadrature operator $\hat q=(\hat a+\hat a^\dag)/\sqrt 2$ and its conjugate momentum, $\hat p=(\hat a-\hat a^\dag)/[i\sqrt{2}]$, such that these operators obey the fundamental commutator relation $[\hat q,\hat p]=i$.
    Now, for the displaced thermal state under study, we find the following expectation values, variances, and covariances:
	\begin{subequations}
	\begin{eqnarray}
		\langle \hat q\rangle &=& \sqrt{2}\mathrm{Re}(\alpha_0),
		\label{eq:q-a-relation}
		\\
		\langle \hat p\rangle &=& \sqrt{2}\mathrm{Im}(\alpha_0),
		\\
		\langle \hat n\rangle &=& \bar n+|\alpha_0|^2,
		\\
		\langle (\Delta \hat q)^2\rangle = \langle (\Delta \hat p)^2\rangle &=& \bar n+\frac{1}{2},
		\\
		\frac{1}{2}\langle \{\Delta \hat q,\Delta \hat p\}\rangle &=& 0,
	\end{eqnarray}
	\end{subequations}
    where we use the anticommutator $\{\,\cdot\,,\,\cdot\,\}$ in the last line for a symmetric ordering.

    Furthermore, we can directly compute the terms that are required to identify the coherence function $\mathcal C$ in Eq. \eqref{eq:QCoherence}.
    In particular, we obtain the purities
	\begin{equation}
		\mathrm{tr}\left(\hat\rho^2\right)
		=\frac{1}{(\bar n+1)^2-\bar n^2}
	\end{equation}
	and
	\begin{equation}
		\mathrm{tr}\left(\hat\rho_{\mathrm{inc.}}^2\right)
		=\frac{\exp\left[
			-\frac{2|\alpha_0|^2}{(\bar n+1)^2-\bar n^2}
		\right]}{(\bar n+1)^2-\bar n^2}
		I_0\left[\frac{2|\alpha_0|^2}{(\bar n+1)^2-\bar n^2}\right].
	\end{equation}
	The difference of both purities then yields Eq. \eqref{eq:Coherence}.
	See also Fig. \ref{fig:QuantCohDTS} for a visualization of $\mathcal C(\hat\rho)$ for the displaced thermal states under study.

%%%%%%%%%%%%%%%%%%%%%%%%%%%%%%%%%%%%%%%%%%%%%%%%%%%%
\section{Monotonicity and entanglement}\label{app:EntanglementMonotonocity}

    Here, we show the expected monotonic behavior of this class of states using the Hilbert-Schmidt inner product for the coherence analysis.
    In addition, known relations between quantum coherence and entanglement are briefly discussed.

    For the class of displaced thermal states that are described by Gaussian phase-space functions, the coherence as measured by the Hilbert-Schmidt distance is well defined and satisfies the expected monotonicity relations.
    That is, the finite expression in Eq. \eqref{eq:Coherence}---see also Fig. \ref{fig:QuantCohDTS}---increases with increasing coherent amplitude $\alpha_0$ and decreases with increasing thermal background $\bar n$.
    Specifically, we obtain
    \begin{equation}
        \frac{\partial \mathcal C}{\partial |\alpha_0|}
        =\frac{4|\alpha_0|e^{-X}}{(2\bar n+1)^2}
        \int_{0}^{2\pi}\frac{d\varphi}{2\pi}\,(1-\cos\varphi)e^{X\cos\varphi}\geq 0,
    \end{equation}
    using $X=2|\alpha|^2/(2\bar n+1)$ and the definition of the Bessel function $I_0$, and
    \begin{equation}
    \begin{aligned}
        \frac{\partial\mathcal C}{\partial \bar n}
        =& -\frac{2\mathcal C}{2\bar n+1}
        \\
        & -\frac{4|\alpha_0|^2 e^{-X}}{(2\bar n+1)^3}
        \int_{0}^{2\pi}\frac{d\varphi}{2\pi}\,(1-\cos\varphi)e^{X\cos\varphi}
        \leq 0,
    \end{aligned}
    \end{equation}
    resulting in the expected monotonicities.

    Concerning the question of entanglement, it was shown in Ref. \cite{KSP16} that an archetypical incoherent operation that allows for converting coherence into entanglement is the map $|n\rangle\mapsto |n\rangle\otimes|n\rangle$.
    In particular, an incoherent mixture is mapped as $\sum_np_n|n\rangle\langle n|\mapsto \sum_{n}p_n|n\rangle\langle n|\otimes|n\rangle\langle n|$, which is a nonentangled state.
    And a coherent superposition maps as
    \begin{equation}
        \sum_n \psi_n|n\rangle\mapsto \sum_n\psi_n|n\rangle\otimes|n\rangle,
    \end{equation}
    which is entangled with Schmidt coefficients $\psi_n$ that determine the amount of entanglement and that are identical to the amplitudes of the input quantum superposition.
    Specifically, the number of nonzero Schmidt coefficients, i.e., the Schmidt rank, determines the dimensionality of the quantum-correlated bipartite system that can be exploited for high-dimensional quantum communication with an encoding alphabet of the size of the Schmidt rank.

    Physically speaking, the considered incoherent operation is a pair-production process in which the pump $|n\rangle_p$ decays into signal-idler pairs, $|n\rangle_s\otimes|n\rangle_i$.
    For example, a parametric down-conversion with a quantized pump maps $|n\rangle_p\mapsto\sum_{j\leq n}\lambda_j|n-j\rangle_p\otimes|j\rangle_s\otimes|j\rangle_i$ when allowing that only $j$ particles decay;
    a measurement projection of the pump in the vacuum state to ensure full conversion then yields the desired process as $({}_p\langle 0|\otimes\hat 1_s\otimes\hat 1_i)\sum_{j\leq n}\lambda_j|n-j\rangle_p\otimes|j\rangle_s\otimes|j\rangle_i\propto |n\rangle_s\otimes|n\rangle_i$.
    It is also noteworthy that a four-particle process of the form $|2n\rangle_p\mapsto|n\rangle_s\otimes|n\rangle_i$, like in a polariton parametric oscillator, works in a similar manner, however, being restricted to even components of the input coherence; this is particularly not of major concern when an infinite number of Fock states are superimposed, such as for the coherent state, $\sum_{n=0}^\infty e^{-|\alpha_0|^2/2}(\alpha_0^n/n!)|n\rangle$.
    General operations that convert coherence into entanglement are discussed in Ref. \cite{KSP16} in detail.

%%%%%%%%%%%%%%%%%%%%%%%%%%%%%%%%%%%%%%%%%%%%%%%%%%%%
\section{Linear coupling}\label{app:LinCoupling}

    In this Appendix, we briefly study the linear coupling between two bosonic modes, $\hat a$ (light) and $\hat b$ (polariton).
    For this purpose, we suppose a Hamiltonian in the interaction picture that takes the form \cite{VW06}
    \begin{equation}
        \hat H_{\mathrm{int}}=\kappa\hat a^\dag\hat b+\kappa^\ast\hat a\hat b^\dag,
    \end{equation}
    where $\kappa\in\mathbb C$ denotes a coupling constant.
    Then the beam-splitter-like input-output relation for the field operators after an interaction time $\tau$ reads
    \begin{equation}
        \begin{pmatrix}
            \hat a\\\hat b
        \end{pmatrix}
        \mapsto
        \begin{pmatrix}
            t & r \\ -r^\ast & t^\ast
        \end{pmatrix}
        \begin{pmatrix}
            \hat a\\\hat b
        \end{pmatrix},
    \end{equation}
    with the transmittance $t=\cos(\vartheta)$ and the reflectance $r=-i\sin(\vartheta)\kappa/|\kappa|$, where $\vartheta=|\kappa|\tau/\hbar$.

    A coherent input state, $|\alpha,\beta\rangle=e^{-(|\alpha|^2+|\beta|^2)/2}\exp(\alpha\hat a^\dag+\beta\hat b^\dag)|0,0\rangle$, is mapped to a coherent state with correspondingly transformed amplitudes, $|t^\ast\alpha-r\beta,t\beta+r^\ast\alpha\rangle$ \cite{VW06}.
    This allows us to consider the two-field state in Glauber-Sudarshan representation after the interaction time $\tau$.
    Suppose the emitted light field is initially in the vacuum state---i.e., a Dirac delta distribution centered at the origin describes the Glauber-Sudarshan distribution---and the polariton state is given by an arbitrary $P(\beta)$.
    This then results in the following output:
    \begin{equation}
    \begin{aligned}
        &\int d^2\alpha\,d^2\beta\,\delta(\alpha)P(\beta)|\alpha,\beta\rangle\langle\alpha,\beta|
        \\
        \mapsto &
        \int d^2\alpha\,d^2\beta\,\delta(\alpha)P(\beta)
        \\
        &\times |t^\ast\alpha-r\beta,t\beta+r^\ast\alpha\rangle\langle t^\ast\alpha-r\beta,t\beta+r^\ast\alpha|
        \\
        =&\int d^2\beta\,P(\beta) |-r\beta,t\beta\rangle\langle -r\beta,t\beta|.
    \end{aligned}
    \end{equation}
    Since we have no direct access to the polariton system, we trace over the second degree of freedom.
    This yields the remaining optical state
   \begin{equation}
   \begin{aligned}
        \hat\rho
        =&\int d^2\beta\, P(\beta)|-r\beta\rangle\langle -r\beta|
        \\
        =&\int d^2\tilde\alpha\, \frac{P\left(\frac{\tilde\alpha}{-r}\right)}{|r|^2}|\tilde\alpha\rangle\langle \tilde\alpha|
   \end{aligned}
    \end{equation}
    by substituting $\tilde\alpha=-r\beta$.

    Therefore, the output light from linear coupling is given by a rescaled version of the polariton state.
    Note that this procedure also resembles the modeling of a loss channel.
    Consequently, the quantum coherence of the polariton system plus the losslike rescaling is mapped onto quantum coherence of the emitted light field. 

\section{Details on numerical analysis and error estimation}\label{app:NumDetails}

    Numerical results are obtained by solving Eqs. \eqref{eq:GPE_psi} and \eqref{eq:GPE_n} via a fourth-order stochastic Runge-Kutta algorithm \cite{HP06} on a finite, two-dimensional grid in real space with edge lengths $L=230.4\,\mu\mathrm{m}$ and a step size of $\sqrt{\Delta V}=0.9\,\mu\mathrm{m}$, satisfying the TWA validity condition, $\hbar\gamma \gg g/\Delta V$ \cite{WS09}.
    For each expectation value, 300 (below threshold, 200 otherwise) realizations were evolved over a time interval of $4~\mathrm{ns}$ with a fixed time step of $0.04~\mathrm{ps}$.
    Additionally, all expectation values and their statistical errors are steady-state average mean, standard deviation, and propagated error, respectively. 
    If the error is smaller than the symbol size, error bars are omitted.
    No error bars are shown in Fig. \ref{fig:sim1} for better visibility.

    As described in the main text, our experimental data are binned to obtain a histogram, which (when normalized) corresponds to the Husimi $Q$ function.
    For any given point in phase space, the mean value $\bar Q$ is the empirical probability of the corresponding bin and the margin of error is given by the standard error, $\sigma(Q)=[\bar Q(1-\bar Q)/\nu]^{1/2}$ for $\nu$ data points.

    The $Q$ function is fitted to the theory of a displaced thermal state for extracting vital parameters, such as $\bar n$ and $|\alpha_0|^2$, which are then used to compute other properties, here denoted by $z$.
    A Monte Carlo error estimation is used to propagate errors to these final properties.
    For this purpose, a sufficiently large sample of random $Q$ functions, $\{Q_i\}_i$, is produced according to a Gaussian distribution with a mean $\bar Q$ and a standard deviation $\sigma(Q)$ for each point in phase space.
    The determination of parameters $z_i$ is carried out as described initially for each sample element $i$ separately.
    The standard deviation of the resulting parameter set $\{z_i\}_i$ yields the uncertainty of $z$.

%%%%%%%%%%%%%%%%%%%%%%%%%%%%%%%%%%%%%%%%%%%%%%%%%%%%
%%%%%%%%%%%%%%%%%%%%%%%%%%%%%%%%%%%%%%%%%%%%%%%%%%%%
%%%%%%%%%%%%%%%%%%%%%%%%%%%%%%%%%%%%%%%%%%%%%%%%%%%%

\end{document}